\documentclass[a4paper,11pt]{article}
\pdfoutput=1 
\usepackage{jheppub}
\usepackage{wrapfig}

\newcommand{\be}{\begin{equation}}
\newcommand{\ee}{\end{equation}}
\newcommand\m{\mu}

\newcommand\n{\nu}
\renewcommand\r{\rho}

\renewcommand\a{\alpha}
\renewcommand\b{\beta}
\renewcommand\l{\lambda}

\def\e{{\rm e}}
\def\d{\partial}
\newcommand{\bseq}{\begin{subequations}}
\newcommand{\eseq}{\end{subequations}}

\newcommand{\ch}{\mathop{\rm ch}\nolimits}
\newcommand{\sh}{\mathop{\rm sh}\nolimits}

\renewcommand{\Im}{\mathop{\rm Im}\nolimits}
\renewcommand{\Re}{\mathop{\rm Re}\nolimits}

\newcommand{\bra}[1]{\langle #1 |}
\newcommand{\ket}[1]{| #1 \rangle}

\begin{document} 
\begin{flushright}
INR-TH/2015-007\\
CERN-PH-TH/2015-063
\end{flushright}
\vspace{-1.5cm}

\title{Semiclassical $S$--matrix for black holes}

\author[a,b,c]{Fedor Bezrukov}
\author[d]{Dmitry Levkov}
\author[a,e,d]{Sergey Sibiryakov}
\affiliation[a]{Physics Department, CERN, CH-1211 Geneva 23,
  Switzerland} 
\affiliation[b]{Physics Department, University of Connecticut, Storrs,
  CT 06269-3046, USA}
\affiliation[c]{RIKEN-BNL Research Center, Brookhaven National
  Laboratory, Upton, NY 11973, USA}
\affiliation[d]{Institute for Nuclear Research of the Russian Academy
  of Sciences,\\  60-th October Anniversary Prospect 7a, Moscow
  117312, Russia}
\affiliation[e]{FSB/ITP/LPPC \'Ecole Polytechnique F\'ed\'erale de
  Lausanne, CH-1015 Lausanne, Switzerland} 
\emailAdd{fedor.bezrukov@uconn.edu}
\emailAdd{levkov@ms2.inr.ac.ru}
\emailAdd{sergey.sibiryakov@cern.ch}

\abstract{We propose a semiclassical method to calculate
  ${\cal S}$--matrix elements for two--stage
  gravitational 
  transitions involving matter collapse into a black hole and
  evaporation of the latter. The method consistently incorporates
  back--reaction of the collapsing and emitted quanta on the metric. We
  illustrate the method in several 
  toy models describing spherical self--gravitating shells in
  asymptotically flat and AdS space--times. We find that electrically
  neutral shells reflect via the above collapse--evaporation process
  with probability $\exp(-B)$, where $B$ is the
  Bekenstein--Hawking entropy
  of the intermediate black hole. This
  is consistent with interpretation of $\exp(B)$ as the
  number of black hole states. 
The same expression for the probability is obtained in the case of charged shells if
one takes into account instability of the Cauchy horizon of the intermediate
Reissner--Nordstr\"om black hole. 
Our semiclassical method opens a new systematic approach to the 
gravitational ${\cal S}$--matrix in the non--perturbative regime.
}

\maketitle

\flushbottom

%%%%%%%%%%%%%%%%%%%%%%%%%%%%%%%%%%%%%%%%%%%%%%%%%%%%%%%%%%%%%%%%%%%%%%%
\section{Introduction}
\label{sec:intro}

Gravitational scattering has been a subject of intensive research over
several decades, see~\cite{Giddings:2011xs} and references
therein. Besides being of its own value, this study presents an
important step towards resolution of the information paradox ---
an apparent clash between unitarity of quantum evolution and 
black hole thermodynamics~\cite{Hawking:1974sw,Hawking:1976ra}.
Recently the interest in this problem has been spurred by the AMPS
(or ``firewall'') argument~\cite{Almheiri:2012rt,Almheiri:2013hfa} 
which suggests that
reconciliation of the black hole evaporation with unitarity would
require drastic departures from the classical geometry in the vicinity
of an old
black hole horizon (see~\cite{Braunstein:2009my1,
  Mathur:2009hf, Braunstein:2009my} for related works). Even 
the minimal versions of such departures appear
to be at odds with the equivalence principle. This calls for putting
all steps in the logic leading to this result on a firmer
footing.

\begin{sloppy}

Unitarity of quantum gravity is strongly supported by the arguments based
on AdS/CFT correspondence \cite{Maldacena:1997re,Maldacena:2001kr}. 
This reasoning is, however, indirect 
and one would like to develop
an explicit framework for testing 
unitarity of black hole evaporation. In particular, one would like to
see how 
the self--consistent quantum evolution
leads to the 
thermal properties of the Hawking radiation 
and 
to test the
hypothesis~\cite{Page:1993wv} that the information about the initial
state producing the black hole is imprinted in subtle correlations
between
the
Hawking quanta. 

\end{sloppy}

\begin{figure}[t]
\centerline{\includegraphics[width=0.85\textwidth]{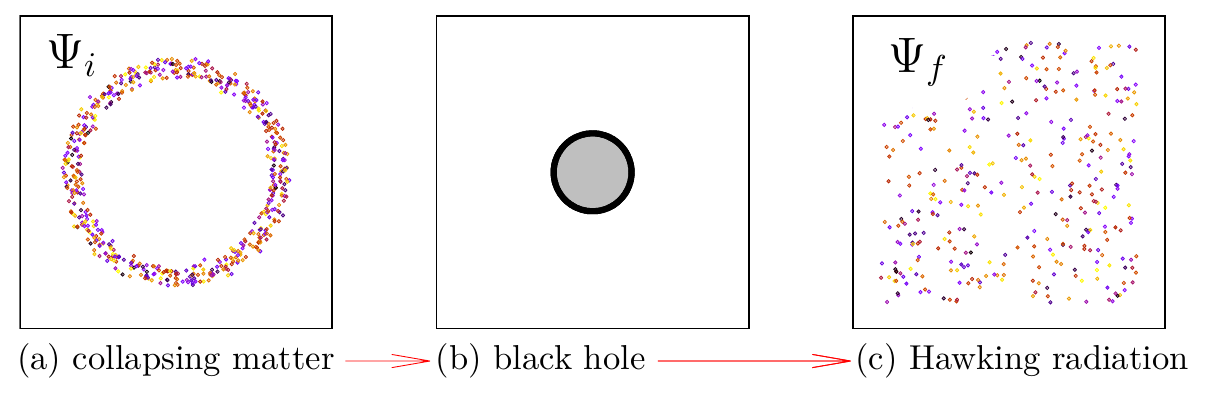}}
\caption{Complete gravitational transition
  involving formation and evaporation of a black hole.\label{fig:collapse}}
\end{figure}

A natural approach is to view the formation of a black hole and its
  evaporation as a two--stage scattering transition, see Fig.~\ref{fig:collapse}.
The initial and
final states $\Psi_i$ and $\Psi_f$ of this process represent free
matter particles and free Hawking quanta in flat space--time. 
They are the asymptotic states of quantum gravity
related by an ${\cal
  S}$--matrix~\cite{Page:1979tc,'tHooft:1996tq,Giddings:2011xs}. 
Importantly, the black hole itself, being metastable, does not
correspond to an asymtotic state.
The ${\cal S}$--matrix
is unitary if black hole formation does not lead to 
information loss. 
The importance of collapse stage for addressing the information
paradox was emphasized in~\cite{Alberghi:2001cm,
  Vachaspati:2006ki, Vachaspati:2007hr, Vachaspati:2007ur,
   Brustein:2013qma}. 

However, calculation of the scattering amplitude for the process in
  Fig.~\ref{fig:collapse} encounters a formidable obstacle:
  gravitational interaction becomes strong in the regime of interest
  and the standard perturbative methods break down
  \cite{Amati:2007ak,Giddings:2009gj}. General
  considerations~\cite{Giddings:2009gj} supported by the perturbative
  calculations~\cite{Amati:2007ak} show that   
scattering of two trans--Planckian particles is accompanied by
an increasingly intensive emission of soft quanta as the threshold of
black hole formation is approached. While this is consistent with the
qualitative properties of the Hawking radiation dominated
 by many quanta, a detailed comparison is not available. A new perturbative
scheme adapted to processes with many particles in the final state has
been recently proposed in~\cite{Dvali:2014ila}; however, its domain
of applicability is yet to be understood.

In this paper we follow a different route. We propose to focus on
  {\em exclusive} processes where both initial and final states
  contain a large number of soft particles. Specifically, one can take
  $\Psi_i$ and $\Psi_f$ to be coherent states with large occupation
  numbers corresponding to the semiclassical wavepackets.
We assume that  
the total energy of the process exceeds the Planck scale, so
that the intermediate black hole has mass well above Planckian.
Then
the overall process is 
expected to be described within the low--energy gravity.
Its amplitude can be evaluated using the semiclassical
methods and will yield the black hole ${\cal S}$--matrix in the
coherent--state
basis~\cite{Tinyakov:1992dr}. A priori, we cannot claim to describe
semiclassically the dominant scattering channel with
Hawking-like final state which is characterized
by low occupation numbers\footnote{It is 
conceivable, however,
that 
  this dominant amplitude can be obtained with a suitable
  limiting procedure. A hint comes from
field theory in flat space where the
cross section of the process $2\to$ {\it many} is recovered from the limit
of the cross section {\it many} $\to$ {\it many} 
\cite{Rubakov:1992ec,Tinyakov:1991fn,Mueller:1992sc}.}. 
Still, it seems a safe bet to expect that, within its domain of
validity\footnote{In line with the common practice, the applicability of the
  semiclassical method will be verified a
  posteriori by subjecting the results to various
  consistency checks.}, this approach will provide valuable information on
the black hole--mediated amplitudes.

A crucial point in the application of semiclassical methods is the
  correct choice of the semiclassical solutions.
Consider the 
amplitude 
\begin{equation}
\label{eq:1}
\langle \Psi_f| \hat{\cal S} | \Psi_i \rangle =
\int {\cal D}\Phi_i {\cal D}\Phi_f
\;\Psi_f^*[\Phi_f]\;\Psi_i[\Phi_i]
\int {\cal D}\Phi
  \,  \mathrm{e}^{iS[\Phi]/\hbar}
\end{equation} 
of transition between the initial and final asymptotic states with
wave functionals $\Psi_i[\Phi_i]$ and $\Psi_f[\Phi_f]$. The path
integral in Eq.~(\ref{eq:1}) runs over all fields $\Phi$ of the theory
including matter  
fields, metrics and ghosts from gauge-fixing of the diffeomorphism
invariance\footnote{The precise definition of the functional
    measure is not required in the leading semiclassical
    approximation, which we will focus on.}; $S$ is the action. 
In the asymptotic past and
future the configurations $\Phi$ in Eq.~(\ref{eq:1}) must describe 
collections of free particles in flat space--time. 
In the semiclassical approach one evaluates the path
integral~(\ref{eq:1}) in the saddle--point approximation.
The saddle--point configuration must inherit the correct 
flat--space asymptotics and, in addition, extremize $S$
  i.e.\ solve the classical equations of motion. However, a naive
  choice of the solution fails to satisfy the former
  requirement. Take, for instance, the solution $\Phi_{cl}$ describing
the {\em classical} collapse. It starts from collapsing 
particles in flat space--time, but 
arrives to a black hole in the asymptotic future. 
It misses the second stage of the scattering process ---
the black hole decay. Thus, it is not admissible as the saddle point of
(\ref{eq:1}). One faces the task of enforcing the correct
asymptotics on the saddle--point configurations.

Furthermore, the saddle--point solutions describing exclusive
transitions are generically
complex--valued and should be considered in 
complexified
space--time~\cite{Tinyakov:1992dr,Rubakov:1992ec}.\footnote{This
  property is characteristic of dynamical tunneling 
  phenomena which have been extensively studied in quantum
mechanics with multiple degrees of freedom~\cite{Creagh}.} 
Even subject to
appropriate boundary conditions, such solutions are
typically not unique.
 Not all of them are relevant: some describe
subdominant processes,
some other, when substituted into the semiclassical expression for the amplitude,
give nonsensical results implying exponentially
large scattering probability. Choosing the dominant physical solution
 presents a
non--trivial challenge.

The method to overcome the two above problems has been developed in 
Refs.~\cite{Bezrukov:2003yf,Bezrukov:2003tg,
Levkov:2007ce,Levkov:2007yn, Levkov:2008xx, Levkov:2009xx}
in the context of scattering in quantum mechanics with multiple
degrees of freedom; it was applied to field theory 
in~\cite{Bezrukov:2003er, Demidov:2011dk}. 
In this paper we adapt this method to the case of gravitational
  scattering.

It is worth emphasizing the important difference between our approach
and perturbative expansion in the classical black hole (or collapsing)
geometry, which is often identified with the semiclassical approximation in 
black hole
physics. 
In the latter case the evaporation is accounted for only
at the one--loop level. This approach is likely to suffer from 
 ambiguities associated with the separation of the system into
 a classical background and quantum fluctuations.
Instead, in our method the semiclassical solutions
by construction encapsulate black hole decay in the
leading order of the semiclassical expansion. They consistently take
into account backreaction of the collapsing
  and emitted matter quanta on the metric. 
Besides, we will find that the 
  solutions describing the process of Fig.~\ref{fig:collapse} are
  complex--valued. They
bypass, via the complexified evolution,
the high--curvature region near the singularity of the 
intermediate black hole. Thus, one does 
 not encounter the problem of resolving the singularity.

One the other hand, the complex--valued saddle--point configurations 
 do not admit a
  straightforward interpretation as classical geometries. In
  particular, they are meaningless for an observer falling into the
  black hole: the latter measures 
local correlation functions given by the path integrals in the {\it
  in--in} formalism --- with different boundary conditions and different
saddle--point configurations as compared to those in
Eq.~(\ref{eq:1}). This distinction lies at the heart of the black hole  
complementarity principle~\cite{Susskind:1993if}. 

Our approach can be applied to any
gravitational system with no symmetry restrictions. However, the 
task of solving nonlinear saddle--point equations is
rather challenging. In this paper we illustrate the method in several
exactly tractable toy models describing spherical gravitating
dust shells. We consider neutral and charged shells in asymptotically flat and anti--de Sitter (AdS)
space--times. Applications to field theory
that are of primary interest are postponed to future.

Although the shell models involve only one collective degree of
freedom ---  the shell radius --- they are believed to capture some
important features of quantum gravity~\cite{Berezin:1997fn,
  Berezin:1999xa, Kraus:1994by, Parikh:1999mf}. Indeed, one can
crudely regard thin shells as narrow wavepackets of an underlying
field theory. In Refs.~\cite{Berezin:1999nn,  Parikh:1999mf,
  Hemming:2000as} emission of Hawking quanta by a black hole is
modeled as tunneling of spherical shells from under the horizon. The
respective emission probability includes back--reaction of the shell on
geometry, 
\begin{equation}
\label{eq:PW}
{\cal P} \simeq \e^{-(B_i-B_f)}\;,
\end{equation}
where $B_i$ and $B_f$ are the Bekenstein--Hawking entropies of the
black hole before and after the emission. It has been argued in
\cite{Parikh:2004ih} that this formula is consistent with
unitary evolution. 

In the context of shell models we consider scattering processes
similar to those in Fig.~\ref{fig:collapse}: a classical contracting
shell forms a black hole and the latter completely decays due to quantum
fluctuations into an  
expanding shell. The initial and final states $\Psi_i$ and  $\Psi_f$
of the process describe free shells in flat or AdS space--times. Our
result for the semiclassical amplitude (\ref{eq:1}) has the 
form
\begin{equation}
\label{eq:2}
\langle \Psi_f | \hat{\cal S} | \Psi_i \rangle \simeq
\,\mathrm{e}^{iS_{reg}/\hbar}\;.
\end{equation}
We stress that it includes backreaction effects.
The probability of transition is 
$$
{\cal P}_{fi}
\simeq\e^{-2\mathrm{Im}\, S_{reg}/\hbar}\;.
$$
We show that for neutral shells it reproduces
Eq.~(\ref{eq:PW}) with $B_i$ equal to the entropy of the
intermediate black hole and $B_f = 0$. This probability is
  exponentially small at $M\gg 1$ when the semiclassical
  approximation is valid. It is consistent 
with the result of Refs.~\cite{Berezin:1997fn, Berezin:1999xa,
  Kraus:1994by, Parikh:1999mf} since the first stage 
of the process, i.e.\ formation of the intermediate black hole, proceeds
classically. 

For charged black holes the same result is recovered once
we take into account instability of the inner Cauchy horizon of
the intermediate Reissner--Nordstr\"om black hole
\cite{Novikov:1980ni, Poisson:1990eh, 
  Ori:1991zz,Brady:1995ni, Hod:1998gy, Frolov:2005ps, Frolov:2006is}.
Our results are therefore consistent with the interpretation of Hawking radiation as
tunneling. However, we obtain important additional information: 
the phases of the ${\cal S}$--matrix elements
which explicitly depend, besides the properties of the intermediate
black hole, on the initial and final states of the process. 

The paper is organized as follows. In Sec.~\ref{sec:method} we
introduce general semiclassical method to compute ${\cal S}$--matrix
elements for scattering via black hole formation and  
evaporation. In Sec.~\ref{sec:shell-reflections} we apply the method
to transitions of 
a neutral shell in asymptotically flat space--time. We also discuss 
relation of the scattering processes to the standard thermal radiation of a
black hole. This analysis is generalized in
Sec.~\ref{sec:massless-shell-ads_4} to a neutral shell in asymptotically
AdS space--time where scattering of the shell admits an AdS/CFT
interpretation. A model with 
an electrically charged shell is studied in
Sec.~\ref{sec:charged-shells}. Section~\ref{sec:concl-prosp} is
devoted to conclusions and discussion of future directions. Appendices
contain technical details.

%\footnote{This approach
%   is similar to the method of constrained instantons \cite{Affleck:1980mp}.}.

%%%%%%%%%%%%%%%%%%%%%%%%%%%%%%%%%%%%%%%%%%%%%%%%%%%%%%%%%%%%%%%%%%%%%%%
\section{Modified semiclassical method}
\label{sec:method}

%%%%%%%%%%%%%%%%%%%%%%%%%%%%%%%%%%%%%%%%%%%%%%%%%%%%%%%%%%%%%%%%%%%%%%%
\subsection{Semiclassical ${\cal S}$--matrix for gravitational scattering}
\label{sec:eval-path-integr}

The ${\cal S}$--matrix is defined as 
\begin{equation}
\label{eq:3}
\bra{\Psi_f}\hat{\cal S}\ket{\Psi_i}
 = \mathrm{lim}_{\tiny \begin{array}{l} t_i\to
    -\infty\\ t_f \to +\infty\end{array}}
\bra{\Psi_f}\hat{U}_0(0,t_f) \, \hat{U}(t_f,t_i)\, 
\hat{U}_0(t_i,0)\ket{\Psi_i}\;,
\end{equation}
where $\hat U$ is the evolution operator;  free evolution
operators $\hat U_0$ on both sides transform from Schr\"odinger to 
the interaction picture. In our case $\hat{U}$ describes quantum
transition in Fig.~\ref{fig:collapse}, while $\hat{U}_0$ generates 
evolution of free matter particles and Hawking quanta in the initial
and final states. The time variable $t\in [t_i,\, t_f]$ is chosen to
coincide with the time of an asymptotic observer at infinity. 

\begin{figure}[t]
\centerline{\includegraphics[width=10cm]{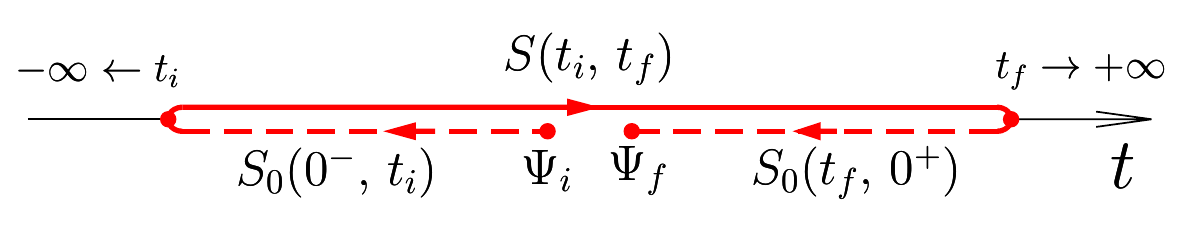}}
\caption{The contour used in the calculation of the ${\cal S}$--matrix
  elements. Quantum transition from $t_i$ to $t_f$ is 
preceded and followed by the free evolution.\label{fig:contour1}}
\end{figure}

Using the path integrals for the evolution operators and taking their
convolutions with the wave functionals of the initial and final states,
one obtains the path integral representation for the 
amplitude\footnote{From now on we work in the 
Planck units $\hbar = c  = G_N = k_B=1$.}  (\ref{eq:3}),
\begin{equation}
\label{eq:4}
\langle \Psi_f | \hat{\cal S}|\Psi_i\rangle = 
\int {\cal D}\Phi \, 
\mathrm{e}^{iS(t_i,\, t_f) + iS_0(0^-,\, t_i) + iS_0(t_f,\, 0^+)} \,
\Psi_i[\Phi_-] \, 
\Psi_f^*[\Phi_+]
\equiv\int {\cal D}\Phi\,\e^{iS_{tot}[\Phi]}\;,
\end{equation}
where $\Phi = \{\phi,\, g_{\mu\nu}\}$ collectively denotes matter and
gravitational fields along the time contour in
Fig.~\ref{fig:contour1}. The quantum measure
${\cal D} \Phi$ should include some ultraviolet regularization as well
as gauge--fixing of the diff--invariance and respective ghosts. 
A non-perturbative definition of this measure 
presents a well--known challenge. Fortunately, the details 
of ${\cal D}\Phi$
are irrelevant
 for our leading--order semiclassical calculations. The
interacting and free actions $S$ and $S_0$ describe evolution along
different parts of the contour. The initial-- and final--state wave functionals
$\Psi_i$ and $\Psi_f$ depend on the fields 
$\Phi_{\mp}\equiv\Phi(t = 0^{\mp})$ at the endpoints of the
contour. In the second equality of Eq.~(\ref{eq:4}) we combined all
factors in the integrand into the ``total action''
$S_{tot}[\Phi]$. Below we mostly focus on nonlinear evolution from
$t_i$ to $t_f$ and take into account contributions from the dashed
parts of the contour in Fig.~\ref{fig:contour1} at the end of the
calculation. 

To distinguish between different scattering regimes,
we introduce a parameter $P$ characterizing the initial
state~\cite{Gundlach:2007gc} ---  say, its average energy. If $P$ is
small, the gravitational interaction is weak and the particles
scatter trivially without forming a black hole. In this regime the
integral in Eq.~(\ref{eq:4}) is saturated by the  saddle--point
configuration $\Phi_{cl}$ satisfying the classical field equations
with boundary conditions related to the 
initial and final states~\cite{Tinyakov:1992dr}. However, if $P$
exceeds a certain critical value $P_*$, the classical solution
$\Phi_{cl}$ corresponds to formation of a black hole. It therefore
fails to interpolate towards the asymptotic out--state $\Psi_f$
living in flat space--time. This marks a breakdown of the standard
semiclassical method for the amplitude (\ref{eq:4}).   

To deal with this obstacle, we introduce a constraint in the path integral
which explicitly guarantees that all field configurations $\Phi$ from 
the integration domain  have flat space--time 
asymptotics~\cite{Levkov:2007yn,Levkov:2008xx,Levkov:2009xx}. 
Namely, we
introduce a functional $T_{int}[\Phi]$ with the following properties: it is
(i)~diff--invariant; (ii)~positive--definite if $\Phi$ is real;
(iii)~finite if $\Phi$ approaches flat space--time at $t\to \pm
\infty$; (iv)~divergent for any configuration containing a black
hole in the asymptotic future. Roughly speaking, $T_{int}[\Phi]$ measures the
 ``lifetime'' of a black hole in the configuration $\Phi$. Possible
choices of this functional will be discussed in the next subsection;
for now let us assume that it exists. Then we consider the identity
\begin{equation}
\label{eq:5}
1 = \int\limits_{0}^{+\infty} dT_0 \, \delta(T_{int}[\Phi] - T_0) =
\int\limits_{0}^{+\infty} dT_0 \int\limits_{-i\infty}^{i\infty}
\frac{d\epsilon}{2\pi i} \, \mathrm{e}^{\epsilon (T_0 -
  T_{int}[\Phi])}\;,
\end{equation}
where in the second equality we used the Fourier representation of the
$\delta$--function. Inserting Eq.~(\ref{eq:5}) into the
integral~(\ref{eq:4}) and changing the order of integration, we
obtain,  
\begin{equation}
\label{eq:6}
\langle \Psi_f | \hat{\cal S} | \Psi_i \rangle =
\int \frac{dT_0 d\epsilon}{2\pi i} \; \mathrm{e}^{\epsilon T_0} \int
     {\cal D}\Phi \, \mathrm{e}^{i(S_{tot}[\Phi] + i\epsilon
       T_{int}[\Phi])}\;.
\end{equation}
The inner integral over $\Phi$ in Eq.~(\ref{eq:6}) has the same form as the
original path integral, but with the modified action
\begin{equation}
\label{eq:7}
S_\epsilon[\Phi] \equiv S_{tot}[\Phi] + i\epsilon T_{int}[\Phi]\;.
\end{equation}
By construction, this integral is restricted to configurations $\Phi$
with $T_{int}[\Phi] = T_0 < \infty$, i.e.\ the ones approaching flat
space--time 
in the asymptotic past and future. In what follows we make this property
explicit by deforming the contour of $\epsilon$--integration to
$\mathrm{Re}\,  \epsilon> 0$. Then configurations with black holes in
the final state i.e.\ with $T_{int}[\Phi] = +\infty$, do not
contribute into the integral at all. 

By now, we have identically rewritten the integral (\ref{eq:4})
  in the form~(\ref{eq:6}). Its value clearly does not depend on the
  form of the regulating functional~$T_{int}[\Phi]$.

Now the amplitude~(\ref{eq:6}) is computed by evaluating the integrals 
over $\Phi$, $\epsilon$ and $T_0$ one after another
in the
saddle--point approximation. The saddle--point configuration
$\Phi_\epsilon$ of the inner integral extremizes the 
modified action (\ref{eq:7}), while the saddle--point equation for
$\epsilon$ gives the constraint
\be
T_{int}[\Phi_\epsilon]  = T_0\;.
\ee
This implies that $\Phi_\epsilon$ has correct
flat--space asymptotics. The integral over $T_0$ is saturated at
$\epsilon=0$. Importantly, we do not substitute $\epsilon=0$ into the saddle--point
equations for $\Phi_\epsilon$, since in that case we would recover the original
classical equations together with  incorrect asymptotics of the saddle--point
solutions. Instead, we understand this equation as the limit
\begin{equation}
\epsilon \to +0\;
\end{equation}
that must be taken at the last stage of the calculation. 
The condition $\Re\epsilon > 0$ is required for convergence of
the path integral (\ref{eq:6}). We obtain the saddle--point
expression (\ref{eq:2}) for the amplitude with the
exponent\footnote{Below we consider only the leading semiclassical exponent. The prefactor in the modified semiclassical approach
  was discussed in \cite{Levkov:2007yn, Levkov:2008xx, Levkov:2009xx}.} 
\begin{equation}
\label{eq:8}
S_{reg} = \lim_{\epsilon \to +0} S_{tot}[\Phi_\epsilon]\;,
\end{equation}
where the limit is taken in the end of the calculation.

Our modified semiclassical method addresses another problem 
mentioned
in the Introduction: it allows to select the relevant saddle--point
configurations from the discrete set of complex--valued solutions to the
semiclassical equations. As discussed in
\cite{Bezrukov:2003yf,Bezrukov:2003tg,Levkov:2007ce}, at $\epsilon > 0$ 
the physical solutions can be obtained from the real classical ones
by analytic continuation in
the parameters of the in-- and out--states. This suggests the following
strategy.  
We pick up a real classical solution $\Phi_0$ describing scattering at a
small value of the parameter ${P<P_*}$. By construction, $\Phi_0$
approaches flat space--time at $t\to \mp\infty$ and gives the 
dominant contribution to the integral (\ref{eq:6}). Next, we modify
the action and gradually increase $\epsilon$ from $\epsilon=0$ to
the positive values constructing a continuous branch of modified solutions
$\Phi_\epsilon$. At $\epsilon\to +0$ these solutions reduce to $\Phi_0$ and
therefore saturate the integral (\ref{eq:6}). We increase the
value of $P$ to $P>P_*$ assuming that continuously deformed
saddle--point configurations $\Phi_\epsilon$ remain
physical\footnote{In other words, we assume that no Stokes
  lines~\cite{Berry:1972na} are crossed in the course of
  deformation. This conjecture has been verified in multidimensional
  quantum mechanics by direct comparison of semiclassical and exact 
  results \cite{Bonini:1999kj,Bezrukov:2003yf,
    Bezrukov:2003tg,Levkov:2007ce, Levkov:2007yn, Levkov:2008xx,
    Levkov:2009xx}.}.
In this way we obtain the modified solutions and the semiclassical
amplitude at any $P$. We stress that our continuation procedure cannot
be performed with the original classical solutions which, 
if continued to $P>P_*$, describe formation of black holes. 
On the contrary, the modified solutions $\Phi_\epsilon$ interpolate
between the flat--space asymptotics at any $P$. They are notably
different from the real classical solutions at $P>P_*$.
At the last step one evaluates the action $S_{tot}$ on
the modified solutions and sends $\epsilon \to +0$ obtaining the leading
semiclassical exponent of the ${\cal S}$--matrix
element, see Eqs.~(\ref{eq:2}), (\ref{eq:8}).

%%%%%%%%%%%%%%%%%%%%%%%%%%%%%%%%%%%%%%%%%%%%%%%%%%%%%%%%%%%%%%%%%%%%%%
\subsection{The functional $T_{int}[\Phi]$}
\label{sec:function-t_int}
Let us construct the appropriate functional $T_{int}[\Phi]$. 
This is particularly simple in the case of reduced models 
with spherically--symmetric
gravitational and matter fields. The general spherically--symmetric
metric has the form
\begin{equation}
\label{eq:9}
ds^2 = \mathrm{g}_{ab}(y) dy^a dy^b + r^2(y) d\Omega^2\;,
\end{equation}
where $d\Omega$ is the line element on a unit two--sphere and
$\mathrm {g}_{ab}$ is the metric in the transverse two--dimensional
space\footnote{We use the signature $(-,+,\ldots)$ for the metrics
  $g_{\m\n}$ and $\mathrm{g}_{ab}$. The Greek indices 
  $\m,\n,\ldots$ are used for the four--dimensional tensors, while the
  Latin ones $a,b,\ldots=0,1$ are reserved for the two--dimensional
space of the spherically reduced model.}.
Importantly, the
radius $r(y)$ of the sphere transforms as a scalar under the
diffeomorphisms of the $y$--manifold. Therefore the functional
\begin{equation}
\label{eq:10}
T_{int} = \int d^2 y \, \sqrt{-\mathrm{g}} \, w(r) F\big(\mathrm{g}^{ab}\d_ar\d_br\big) \;, 
\end{equation}
is diff--invariant. Here $w(r)$ and $F(\Delta)$ are non--negative
functions, so that the functional (\ref{eq:10}) is positive-definite.
We further require that $F(\Delta)$
vanishes if and only if $\Delta=1$. Finally, we assume that $w(r)$ significantly
differs from zero only at $r\lesssim r_w$, where $r_w$ is some fixed value, and falls off sufficiently
fast at large $r$. An example of functions satisfying these conditions
is $w(r)=\delta(r-r_w)$, $F(\Delta)=(\Delta-1)^2$.  

To understand the properties of the functional~\eqref{eq:10}, we
consider the Schwarzschild frame where $r$ is the spatial
coordinate and the metric is diagonal. The functional~(\ref{eq:10}) takes
the form, 
\begin{equation}
\label{eq:11}
T_{int} = \int dt dr \sqrt{-\mathrm{g}} \, w(r) F\big(\mathrm{g}^{11}\big)\;.
\end{equation}
Due to fast falloff of $w(r)$ at infinity the integral over $r$ in
this expression is finite. However, convergence of the time
integral depends on the asymptotics of the metrics in the past and
future. In flat space--time $\mathrm{g}^{11}=1$ and the integrand in
Eq.~(\ref{eq:10}) vanishes.
Thus, the integral over $t$ is finite if $\mathrm{g}_{ab}$
approaches the flat metric at $t\to \pm \infty$. Otherwise the
integral diverges. In particular, any classical solution with a black hole in
the final state leads to linear divergence  at
$t\to +\infty$ because the Schwarzschild metric is static and
$\mathrm{g}^{11}\neq 1$. Roughly speaking, $T_{int}$ 
can be regarded as the Schwarzschild
time during which matter fields efficiently interact with gravity
inside the region $r < r_w$. If matter leaves this 
region in finite time, $T_{int}$ takes finite values. It diverges
otherwise. 
Since the 
functional (\ref{eq:10}) is diff--invariant, these
properties do not depend on the particular choice of the coordinate
system. 

The above construction will be sufficient for the purposes of
the present paper. 
Beyond the spherical symmetry one can use the functionals 
$T_{int}[\Phi]$ that involve, e.g., an integral of 
the square of the Riemann tensor, or the Arnowitt--Deser--Misner (ADM) mass
inside a large volume. 
Recall the final result for the
${\cal S}$--matrix does not depend on the precise choice of the
functional $T_{int}$. Of course, this functional should satisfy 
 the conditions (i)-(iv) listed before Eq.~(\ref{eq:5}).

%%%%%%%%%%%%%%%%%%%%%%%%%%%%%%%%%%%%%%%%%%%%%%%%%%%%%%%%%%%%%%%%%%%%%%%
\section{Neutral shell in flat space--time}
\label{sec:shell-reflections}

%%%%%%%%%%%%%%%%%%%%%%%%%%%%%%%%%%%%%%%%%%%%%%%%%%%%%%%%%%%%%%%%%%%%%%
\subsection{The simplest shell model}
\label{sec:models}
We illustrate the method of Sec.~\ref{sec:method} in the spherically symmetric model of
gravity with thin dust shell for matter. The
latter is parameterized by a single  collective coordinate --- the
shell radius  
$r(\tau)$ --- depending on the proper time along the shell $\tau$. This
is a dramatic simplification as 
compared to the realistic case of matter described by dynamical
fields. Still, one can interprete the shell as a toy model for the evolution of narrow wavepackets in field theory. In particular, one expects
that the
shell model captures essential features of gravitational
transition between such wavepackets.\footnote{Note that our approach
  does not require  complete solution of the quantum shell model which
  may be ambiguous. Rather, we look for complex solutions of the
  classical equations saturating the path integral.}

The minimal action for a spherical dust shell is 
 \begin{equation}
 \label{Ssimplest}
 S_{shell}=-m\int d\tau\; 
 \end{equation}
where $m$ is the shell mass. However, such a shell always collapses
into a black hole and hence is not sufficient for our
purposes. Indeed, as explained in Sec.~\ref{sec:eval-path-integr}, in
order to select 
the physically relevant semiclassical solutions we need a parameter
$P$ such that an initially contracting shell reflects classically at
$P<P_*$ and forms a black hole at $P>P_*$.  We therefore generalize the model~\eqref{Ssimplest}. To
this end we assume that the shell is assembled from particles with
nonzero angular momenta. At each point on the shell the velocities
of the constituent particles are uniformly distributed in the tangential
directions, so that the overall configuration is
spherically--symmetric\footnote{Similar models are used in
  astrophysics to describe structure formation \cite{Sikivie:1996nn}.}. 
The corresponding shell action is \cite{Berezin:2000vv}
\begin{equation}
\label{eq:18}
S_{shell}  = - \int  m_{\mathrm{eff}} \, d\tau \;,
\qquad  \qquad m_{\mathrm{eff}}^2 = m^2 + L^2/r^2(\tau)\;,
\end{equation}
where $L$ is a parameter proportional to the angular momentum of the
constituent particles. Its nonzero value provides a centrifugal barrier
reflecting classical shells at low energies. Decreasing this
parameter, we arrive to the regime of classical gravitational
collapse. In what follows we switch between the scattering regimes by
changing the parameter $L\equiv P^{-1}$. For completeness we derive the
action (\ref{eq:18}) in Appendix~\ref{sec:shell-rotating-dust}. 

Gravitational sector of the model is described by the Einstein--Hilbert
action with the Gibbons--Hawking term,
\begin{align}
\label{eq:12}
&S_{EH} = \frac{1}{16\pi} \int_{\cal V} d^4 x \sqrt{-g} \, R\;, \\
\label{eq:14}
&S_{GH} = \frac{1}{8\pi} \int_{\partial {\cal V}} \kappa\,d^3
\sigma\,\sqrt{|h|} \,(K - K_0)\;.
\end{align}
Here the metric $g_{\mu\nu}$ and curvature scalar $R$ are defined
inside the space--time volume ${\cal V}$ with the boundary\footnote{We impose
 the Dirichlet boundary conditions on the variations of $g_{\mu\nu}$ at
 $\partial {\cal V}$.} $\partial {\cal V}$. The latter 
consists of a time--like surface at spatial infinity 
$r = r_\infty \to +\infty$ and space-like surfaces at the initial
and final times $t = t_{i,f} \to \mp \infty$. In Eq.~(\ref{eq:14})
$\sigma$ are the coordinates on the boundary, $h$ is the determinant of
the induced metric, while $K$ is the extrinsic curvature involving the
outer normal. The parameter $\kappa$ 
equals $+1$ ($-1$) at the time--like (space--like) portions of the
boundary. To obtain zero gravitational action in flat space--time, we
subtract the regulator $K_0$ which is equal to the flat--space extrinsic
curvature of the boundary~\cite{Poisson}. For the sphere at infinity
$K_0=2/r_{\infty}$, while the initial-- and final--time hypersurfaces
have $K_0=0$. The Gibbons--Hawking term (\ref{eq:14}) will play an
important role in our analysis.

\begin{figure}[t]
\centerline{
\includegraphics[width=5cm]{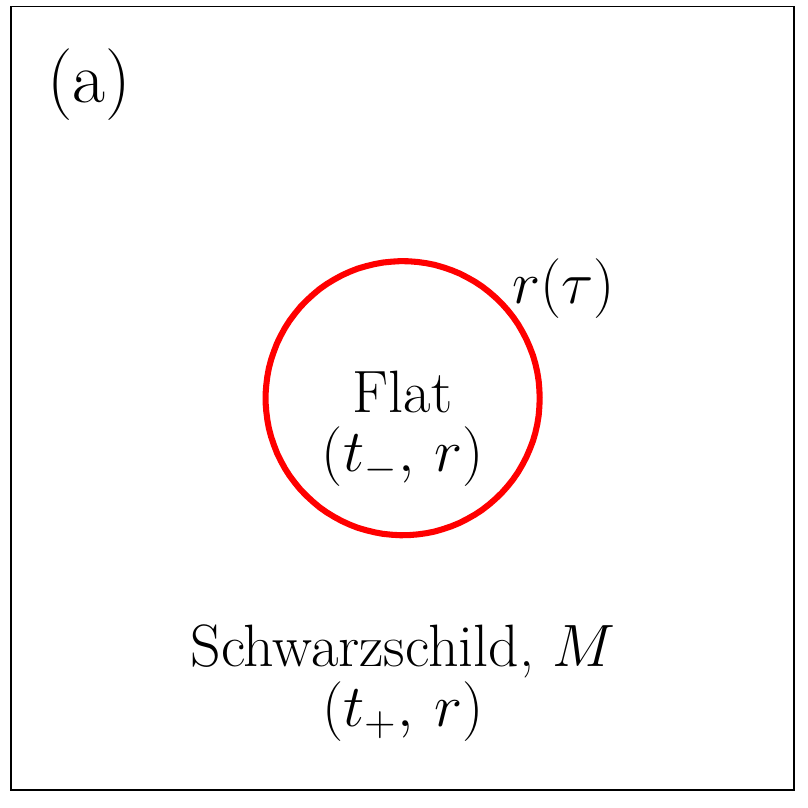}\hspace{2cm}
\includegraphics[width=5cm]{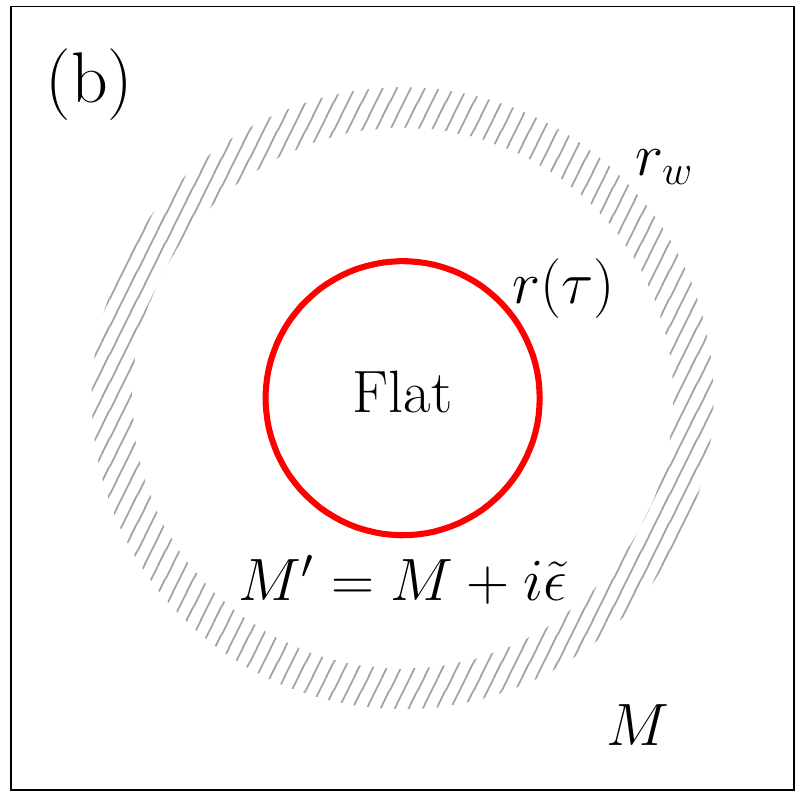}}
\caption{Gravitational field of the spherical dust shell: (a) in the original model, (b) in the model with modification at~$r\approx r_w$.\label{fig:shell}}
\end{figure}

Let us first discuss the classical dynamics of the
system. Equations of motion follow from variation of the total
action
\begin{equation}
\label{eq:31}
S = S_{shell} + S_{EH} + S_{GH}
\end{equation} 
with respect to the metric $g_{\mu\nu}$ and the shell trajectory
$y^a(\tau)$. 
In the regions inside and outside the shell the metric satisfies
vacuum Einstein equations and therefore, due to Birkhoff theorem, is given by
the flat and Schwarzschild solutions, respectively, see  Fig.~\ref{fig:shell}a.
Introducing the spherical  coordinates  
$(t_{-}, \, r)$ inside the shell and Schwarzschild coordinates
$(t_{+},\, r)$ outside, one writes the inner and outer metrics in
the universal form
\begin{equation}
\label{eq:21}
ds^2_{\pm} = -f_{\pm}(r) dt_\pm^2 + \frac{dr^2}{f_\pm(r)} + r^2 d\Omega^2\;,
\end{equation}
where
\begin{equation}
\label{eq:19}
f_{-} = 1\;, \qquad\qquad f_{+} = 1 - 2M/r\;.
\end{equation}
The parameter $M$ is the ADM mass which 
coincides with the total energy of the shell. 
In what follows we
will also use the Schwarzschild radius $r_h \equiv 2M$.
For the validity of the semiclassical
approach we assume that the energy is higher than Planckian, $M\gg 1$.

\begin{figure}[t]
\centerline{\includegraphics[width=0.45\textwidth]{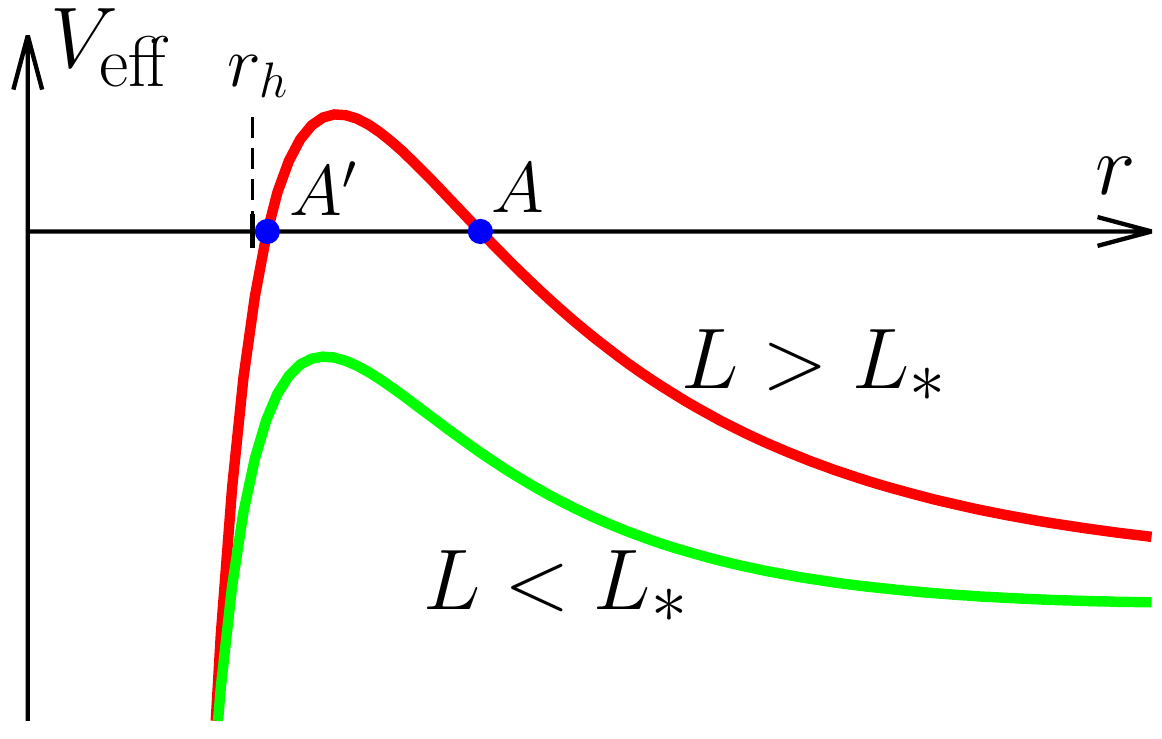}}
\caption{Effective potential for the shell motion.\label{fig:pot}}
\end{figure}

Equation for the shell trajectory is derived in
Appendix~\ref{sec:israel-condition} by matching the inner and outer
metrics at the shell worldsheet with the Israel junction
conditions~\cite{Israel:1966rt,Berezin:1987bc}. It
can be cast into the form
of an equation of motion for a particle with zero energy in an
effective potential,
\begin{align}
\label{eq:30}
&\dot{r}^2 + V_{\mathrm{eff}}(r) = 0\;,\\ 
\label{eq:38}
&V_{\mathrm{eff}}
= f_{-} -\frac{r^2}{4m_{\mathrm{eff}}^2} \left(f_{+}-f_{-} -
\frac{m_{\mathrm{eff}}^2}{r^2}\right)^2\\
&~~~~~=1-\frac{\left(L^2 + m^2r^2 +
  2Mr^3\right)^2}{4r^4(L^2 + m^2 r^2)}\;.
\notag
 \end{align}
 This equation incorporates gravitational effects as well as
the backreaction of the shell on the spacetime metric.
 The potential $V_{\mathrm{eff}}(r)$ goes to $-\infty$ at $r\to 0$ and asymptotes to a
negative value\footnote{Recall that the shell energy $M$ is always
  larger than its rest mass $m$.} 
$1-M^2/m^2$ at $r=+\infty$, see Fig.~\ref{fig:pot}.
At large enough $L$ the potential crosses zero at the points $A$ and
$A'$ --- the turning points of classical motion. A shell coming
from infinity reflects from the point $A$ back to $r=+\infty$. When
$L$ decreases, the turning points approach each 
other and coalesce at a certain critical value\footnote{For a massless shell
  ($m=0$) the critical value is 
  $L_* = 27M^2/8$.} $L=L_*$. At even smaller $L$ the turning points
migrate into the complex plane, see Fig.~\ref{fig:potential} (upper
left panel), and the potential barrier disappears. Now a classical
shell coming from infinity goes all the way to $r=0$. This
is the classical collapse.  

Now, we explicitly see an obstacle for finding the reflected
semiclassical solutions at $L<L_*$ with the method of continuous
deformations. Indeed, at large $L$ the reflected solutions $r=
r(\tau)$ are implicitly defined as 
\begin{equation} 
\label{eq:trajimpl}
\int\limits^{r(\tau)} \frac{dr}{\sqrt{-V_{\mathrm{eff}}(r)}}=\tau\;,
\end{equation}
where the square root is positive at $r\to +\infty+i\, 0$. 
The indefinite integral is performed along the contour ${\cal C}$
running from $r = +\infty - i0$ to $r = +\infty + i0$ and encircling the
turning point $A$~--- the branch point of the integrand (see the upper
left panel of Fig.~\ref{fig:potential}).  As $L$ is lowered, the
branch point moves and the integration contour stays attached to
it. However, at $L=L_*$ when the branch points $A$ and $A'$ coalesce, the
contour ${\cal C}$ is undefined. It is therefore impossible to obtain
reflected semiclassical solutions at $L<L_*$ from the classical solutions
at $L>L_*$.

\subsection{Modification}
\label{sec:modification}

To find physically relevant reflected trajectories at $L<L_*$, we use
the method of Sec.~\ref{sec:method} and add an imaginary term
$i\epsilon T_{int}$ to  the action. We consider
$T_{int}$ of the form (\ref{eq:10}), where the function $w(r)$ is
concentrated in the  vicinity of $r=r_w$. The radius $r_w$ is chosen
to be large enough, in 
particular, larger than  the
Schwarzschild radius $r_h$ and the position $r_A$ of the right turning point
$A$. Then the Einstein equations are 
modified only at $r\approx r_w$, whereas the geometries inside and
outside of this  layer are given by the Schwarzschild solutions with
masses $M'$ and $M$, see Fig.~\ref{fig:shell}b. To connect these
masses, we solve the modified Einstein equations in 
the vicinity of $r_w$. Inserting general spherically symmetric metric
in the Schwarzschild frame,  
\begin{equation}
\label{eq:36}
ds^2 = -f(t,r) dt^2 + \frac{dr^2}{\tilde f(t,r)} + r^2 d\Omega^2\;,
\end{equation}
into the $(tt)$ component of Einstein equations, we obtain,
\begin{equation}
\label{eq:37}
\frac{\partial_r \tilde f}{r} - \frac{1 - \tilde f}{r^2} =
 \frac{2i\epsilon}{r^2} w(r) \, F(\tilde f)\;.
\end{equation} 
The solution reads\footnote{The function $\tilde f$ is
  time--independent due to the $(tr)$ equation.}, 
\begin{equation}
\tilde f = 1 - \frac{2 \tilde{M}(r)}{r} \;, \qquad\mathrm{where}
\qquad \tilde{M}(r) = M + i\epsilon \int_{r}^{+\infty} dr' w(r') F(\tilde f) \;.
\end{equation}
This gives the relation 
\begin{equation}
\label{eq:35}
M' = M + i\tilde{\epsilon}~,
\qquad \qquad \tilde \epsilon=
\epsilon \int\limits_{r\approx r_w}dr'\, w F\;.
\end{equation}
Here $\tilde{\epsilon}>0$ is the new parameter of
modification. 
As before, the ADM mass $M$ of the system is conserved
in the course of the evolution. It coincides with the initial and
final energies of the shell which are, in turn, equal, as will be shown 
  in Sec.~\ref{sec:cal-s-matrix}, to the initial-- and
  final--state energies in the quantum scattering problem. Thus, 
   $M$ is real, while the mass $M'$ of the Schwarzschild
space--time surrounding the shell acquires a positive imaginary
part\footnote{In this setup the method of Sec.~\ref{sec:method} is
  equivalent to analytic continuation of the scattering amplitude into the
  upper half--plane of complex ADM energy, cf.~\cite{Bezrukov:2003tg}.}. 
The shell dynamics in this case is still described by Eq.~(\ref{eq:30}),
where $M$ is replaced by $M'$  in the potential
(\ref{eq:38}). Below we find semiclassical solutions for small 
$\tilde\epsilon >0$. In the end $\tilde\epsilon$ will be sent to zero.  

\begin{figure}
\centerline{\includegraphics[width=0.8\textwidth]{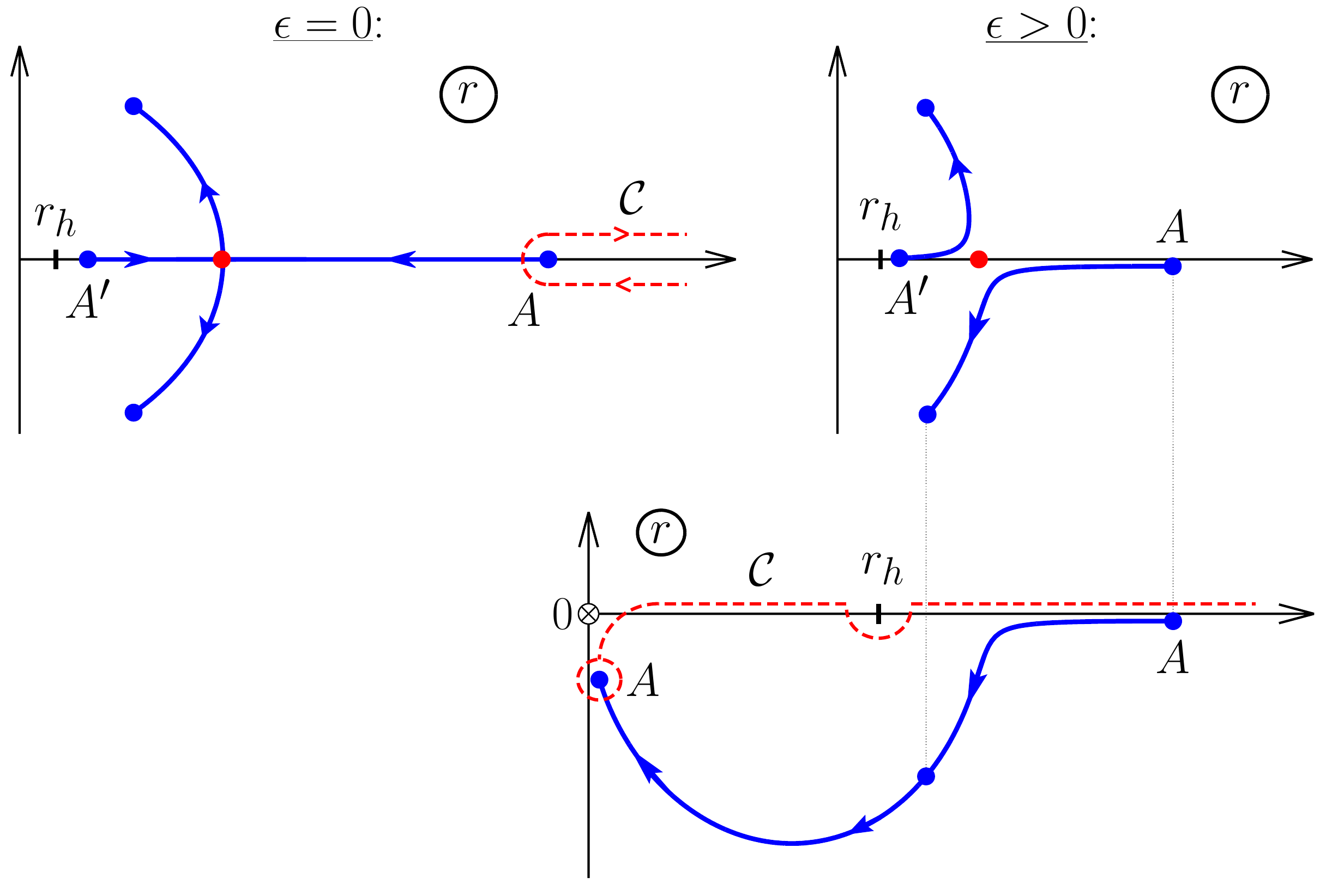}}
\caption{Motion of the turning points in
  the complex $r$--plane. The points move along the arrows as $L$
  decreases. The cases of classical shell
  (upper left panel) and shell with the
modified action (right panels) are shown.  
\label{fig:potential}}
\end{figure}

Let us study the effect of the modification~\eqref{eq:35} on the
semiclassical trajectories $r = r(\tau)$ in
Eq.~(\ref{eq:trajimpl}). At $L>L_*$ the complex terms in $V_\mathrm{eff}$
are negligible and the reflected trajectory is 
obtained with the same contour ${\cal C}$ as before, see 
the upper left panel of Fig.~\ref{fig:potential}. The modification of
$V_{\mathrm{eff}}$ becomes important when $L$ gets close to $L_*$ and
the two turning points $A$ and $A'$ approach each other. Expanding the
potential in the neighborhood of the maximum, we write,
\begin{equation}
\label{eq:40}
V_{\mathrm{eff}}(r) \approx V_{max} -\mu^2 (r - r_{max})^2\;,
\end{equation}
where $V_{max}$, $\mu$ and $r_{max}$ depend on $L$ and $M'$. For real
$M'=M$ the extremal value $V_{max}$ is real and crosses zero when $L$
crosses $L_*$, whereas the parameters $\mu^2>0$ and $r_{max}$ remain approximately
constant. The shift of $M'$ into the upper complex
half--plane gives a negative imaginary part to $V_{max}$,  
\begin{equation}
\label{eq:39}
\mathrm{Im}\, V_{max} = \tilde{\epsilon} \; \frac{\partial
  V_{max}}{\partial M'}  + O(\tilde{\epsilon}^2) < 0\;,
\end{equation}
where the last inequality follows from the explicit
form~(\ref{eq:38}). Now, it is straightforward to track the motion of the
turning points using Eq.~(\ref{eq:40}) as $L$ 
decreases below $L_*$. Namely, $A$ and $A'$ are shifted into the lower
and upper half--planes as 
shown in Fig.~\ref{fig:potential} (upper right panel). Importantly,
these points never coalesce. Physically relevant reflected solution at
$L<L_*$ is obtained by continuously deforming the contour of
integration in 
Eq.~(\ref{eq:trajimpl}) while keeping it attached to the same
turning point. As we anticipated in
Sec.~\ref{sec:method}, a smooth branch of reflected semiclassical
solutions parameterized by $L$ exists in the modified system. 
\begin{sloppy}

If $L$ is slightly smaller than $L_*$, the relevant saddle--point
trajectories reflect at ${\mathrm{Re}\, r_A > r_h}$ and hence never
cross the horizon. A natural interpretation of the corresponding quantum transitions is over--barrier reflection from the centrifugal
potential. However, as $L$ decreases to $L\to 0$, the centrifugal
potential vanishes. One expects that the semiclassical trajectories in
this limit describe complete gravitational transitions 
proceeding via formation and decay of a black hole. 

\end{sloppy}
After establishing the correspondence between the solutions at $L>L_*$
and $L<L_*$, we take\footnote{This may be impossible in more
  complicated systems~\cite{Bezrukov:2003yf,
    Bezrukov:2003tg,Levkov:2007yn, Levkov:2008xx} where the relevant 
  saddle--point trajectories do not exist at $\tilde{\epsilon}=0$. In
  that case one works at nonzero $\tilde{\epsilon}$ till the end of
  the calculation.} $\tilde{\epsilon} = 0$. This leaves us with the
original semiclassical equations which do not include the imaginary 
regularization terms.

We numerically traced the motion of the turning point 
$A$ as $L$ decreases from large to small
values, see Fig.~\ref{fig:potential} (lower
panel). It approaches the
singularity $r=0$ at $L\to 0$.
This behavior is confirmed analytically in Appendix~\ref{app:tp}.
Thus, at small
$L$ the contour ${\cal C}$ goes essentially along the real axis
making only a tiny excursion into the complex plane near the
singularity. It encircles the horizon $r=r_h$ from below. 

One remark is in order. For the validity of the low--energy gravity
the trajectory should stay in the region of sub-Planckian curvature,
$R_{\m\n\l\r}R^{\m\n\l\r}\sim M^2/r^6\ll 1$. This translates into the
requirement for the turning point 
\be
\label{rAM}
|r_A|\gg M^{1/3}\;.
\ee
On
the other hand, we will see shortly that the dependence of 
the semiclassical amplitude on $L$ drops off 
at $L \ll  L_* \sim M^2$; we always understand the limit 
$L\to 0$ in the sense of this inequality.
Using the expressions for $r_A$ from Appendix \ref{app:tp} one 
verifies that the condition
(\ref{rAM}) can be satisfied simultaneously with $L\ll L_*$ in the
semiclassical regime 
$M \gg 1$.

%%%%%%%%%%%%%%%%%%%%%%%%%%%%%%%%%%%%%%%%%%%%%%%%%%%%%%%%%%%%%%%%%%%%%%
\subsection{${\cal S}$--matrix element}
\label{sec:cal-s-matrix}

\paragraph{\em The choice of the time contour.}
 The action $S_{reg}$ entering the amplitude 
(\ref{eq:2}) is computed along the
contour in complex plane of the asymptotic observer's time $t\equiv
t_+$. Since we have already found the physically  
relevant contour ${\cal C}$ for $r(\tau)$,
let us calculate the Schwarzschild time $t_+(r)$ along this contour. We
write,
\begin{equation}
\label{eq:time}
t_+(r)=\int^{r} dr\; \frac{dt_+}{dr} =\int^r\frac{dr}{\sqrt{-V_{\rm
      eff}(r)}}\; 
\frac{\sqrt{f_+(r)-V_{\rm eff}(r)}}{f_+(r)}\;,
\end{equation}
where the indefinite integral runs along ${\cal C}$. In
Eq.~\eqref{eq:time} we used the the definition of the proper time
implying 
\[
f_+\bigg(\frac{dt_+}{dr}\bigg)^2=\frac{1}{\dot r^2}+\frac{1}{f_+}\;,
\]
and expressed $\dot r^2$ from
Eq.~(\ref{eq:30}). The integrand in  Eq.~(\ref{eq:time}) has a pole at
the horizon $r= r_h$,  
$f_+(r_h)=0$, which is encircled from below, see Fig.~\ref{fig:potential}, lower
panel. The half--residue at this pole contributes $i\pi r_h$ to $t_+$
each time the contour ${\cal C}$ passes close to it. The
  contributions have the same sign: although the contour ${\cal C}$
  passes the horizon in the opposite directions, the square
  root in the integrand changes sign after
  encircling 
  the turning point. Additional imaginary
contribution comes from the integral between the real $r$--axis and the
turning point $A$; this contribution vanishes at $L\to 0$. 

\begin{figure}[t]
  \vspace{-5mm}
\centerline{\includegraphics[width=10cm]{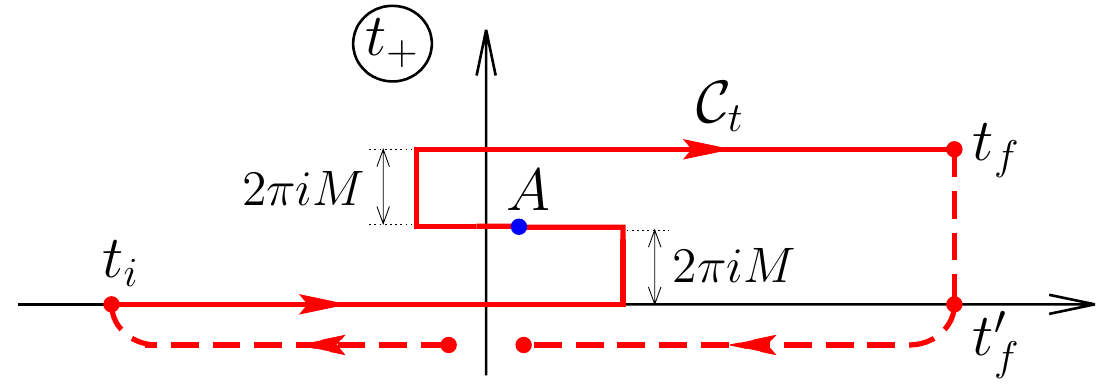}}
\caption{The time contour corresponding to the semiclassical
  solution at small $L$. Solid and dashed lines
  correspond to interacting and free evolution respectively, 
  cf. Fig.~\ref{fig:contour1}. \label{fig:contour2}}
\end{figure}

The image ${\cal C}_t$ of the contour ${\cal C}$ is shown in 
Fig.~\ref{fig:contour2}, solid line. Adding free evolution from $t_+ =
0^-$ to $t_+ = t_i$ and from $t_+ = t_f$ to $t_+ = 0^+$ (dashed
lines), we obtain the contour analogous to the one in
Fig.~\ref{fig:contour1}.
One should not worry about the complex value of $t_f$ in
Fig.~\ref{fig:contour2}: the limit $t_f\to +\infty$ in the definition
of $S$--matrix implies that $S_{reg}$ does not depend on
$t_f$. Besides, the semiclassical solution $r = r(t_+)$ is
an analytic function of $t_+$ and the contour ${\cal C}_t$ can be
deformed in complex plane as long as it does not cross 
the singularities\footnote{In fact, ${\cal C}_t$ is separated from the
  real time axis by a singularity where $r(t_+) = 0$. This is the
  usual situation for tunneling solutions in quantum mechanics and
  field theory~\cite{Bezrukov:2003yf,Bezrukov:2003tg}. Thus, $S_{reg}$
  cannot be computed along the contour in Fig.~\ref{fig:contour1};
  rather, ${\cal C}_t$
  or an equivalent contour should be used.} of $r(t_+)$. Below we calculate the
action along ${\cal C}_t$ because the shell position and the metrics are
real in the initial and final parts of this contour. This simplifies
the calculation of the Gibbons--Hawking terms at $t_+=t_i$ and
$t_+=t_f$.

\paragraph{\em Interacting action.}
Now, we evaluate the action of the interacting system
$S(t_i,t_f)$ entering $S_{reg}$. We rewrite the shell action as 
\be
\label{eq:shellact}
S_{shell}=-\int_{\cal C}\frac{dr}{\sqrt{-V_{\rm eff}}}\,m_{\rm eff}\;,
\ee
where Eq.~\eqref{eq:trajimpl} was taken into account. The
Einstein--Hilbert action (\ref{eq:12}) is simplified using 
the trace of the Einstein equations, $R=-8\pi T_{shell\,
  \m}^{\phantom{shell}\m}$, and the energy--momentum tensor of the shell
computed in Appendix \ref{sec:israel-condition}, Eq.~(\ref{eq:20}), 
\begin{equation}
\label{eq:41}
S_{EH} = \int_{\tau_i}^{\tau_f} d\tau \, \frac{m^2}{2m_{\mathrm{eff}}}
=\int_{\cal C}\frac{dr}{\sqrt{-V_{\rm eff}}}\, \frac{m^2}{2m_{\mathrm{eff}}}\;.
\end{equation}
An important contribution comes from the Gibbons--Hawking
term at spatial infinity $r = r_{\infty} \to 
+\infty$. The extrinsic
curvature reads,  
\begin{equation}
\label{eq:42}
K\Big|_{r_\infty} = \frac{r f_{+}' + 4f_{+}}{2r\sqrt{f_{+}}}
\Bigg|_{r_{\infty}} = \frac{2}{r_{\infty}} - \frac{M}{r_{\infty}^2} +
O(r_{\infty}^{-3}) \;.
\end{equation}
The first term here is canceled by the regulator $K_0$ in
Eq.~(\ref{eq:14}). The remaining 
expression is finite at $r_\infty\to +\infty$,
\begin{equation}
\label{eq:45}
S_{GH}\Big|_{r_\infty} = - \frac{M}{2} \int dt_{+}=
-\frac{M}{2}
\int_{\cal C}\frac{dr}{\sqrt{-V_{\rm eff}}}\, 
\frac{\sqrt{f_+-V_{\rm eff}}}{f_+}\;,
\end{equation}
where we transformed to integral running along the
contour ${\cal C}$ using
Eq.~(\ref{eq:time}). Note that this contribution contains an
imaginary part 
\begin{equation}
\label{eq:ImGH}
\Im S_{GH}\Big|_{r_\infty}=-\frac{M}{2}\Im (t_f-t_i)\;.
\end{equation}
Finally, in Appendix \ref{sec:gibb-hawk-terms}  we evaluate the
Gibbons--Hawking terms at the initial-- and final--time
hypersurfaces. The result is  
\begin{equation}
\label{eq:68}
S_{GH}\Big|_{t_{i,f}} = \frac{\sqrt{M^2-m^2}}{2}\; r_{i,f}  +  
\frac{M(2M^2-m^2)}{4\sqrt{M^2-m^2}}\;,
\end{equation}
where $r_{i,f}$ are the radii of the shell at the endpoints of the
contour ${\cal C}$. The latter radii are real, and so are the terms
(\ref{eq:68}).  

Summing up the above contributions,
one obtains, 
\begin{multline}
\label{eq:Sint}
S(t_i,t_f)=\!\!\int\limits_{\cal C}\!\!\frac{dr}{\sqrt{-V_{\rm eff}}}
\bigg[\frac{m^2-2m^2_{\rm eff}}{2 m_{\rm eff}}-
\frac{M}{2}\frac{\sqrt{f_+-V_{\rm eff}}}{f_+}\bigg]
\\+\frac{\sqrt{M^2\!-\!m^2}}{2}\, (r_i+r_f)  +  
\frac{M(2M^2\!-\! m^2)}{2\sqrt{M^2\! -\! m^2}}\,.
\end{multline}
This expression contains linear and logarithmic divergences
when $r_{i,f}$ are sent to infinity. Note that the divergences appear
only in the real part of the action and thus affect only the phase of
the reflection amplitude but not its absolute value.

\paragraph{\em Initial and final--state contributions.}
The linear divergence in Eq.~(\ref{eq:Sint}) is related to free motion of the shell in the
asymptotic region ${r\to +\infty}$, whereas the logarithmic one
is due to the $1/r$ tails of the gravitational interaction in this
region. Though the $1/r$ terms in the Lagrangian represent
vanishingly small gravitational forces in the initial and final
states, they produce logarithmic divergences in $S(t_i,t_f)$ when
integrated over the shell 
trajectory. To obtain a finite matrix element, we
include\footnote{This relies on the freedom in splitting the total
  Lagrangian into ``free'' and  ``interacting'' parts.}
these terms in the definition of the free action $S_0$. In
Appendix~\ref{app:free} the latter action is computed for the shell with
energy $M$,
\begin{equation}
\label{eq:44}
S_0 (0^-,t_i)= \int_{r_1}^{r_i} p_i(r,M) dr-Mt_i\;,~~~~~~~
S_0 (t_f,0^+)= \int_{r_f}^{r_2} p_f(r,M) dr+Mt_f\;,
\end{equation}
where $r_{1,2}$ are the positions of the shell at $t_+=0^\mp$ and
\begin{equation}
\label{eq:77}
p_{i,f}(r,M) =\mp \bigg[\sqrt{M^2-m^2} 
+ \frac{M(2M^2-m^2)}{2r\sqrt{M^2-m^2}}\bigg]
\end{equation}
are the initial and final shell momenta with  $1/r$
corrections.

The path integral (\ref{eq:4}) for the amplitude involves free wavefunctions $\Psi_i(r_1)$ and $\Psi_f(r_2)$ of the initial and final
states. We consider the semiclassical wavefunctions of the shell with
fixed energy $E$,
\begin{equation}
\label{eq:33}
\Psi_i(r_1) \simeq \exp\left\{i \int_{r_0}^{r_1} p_i(r',E) dr'\right\}\;, \qquad
\qquad \Psi_f(r_2) \simeq \exp\left\{i \int_{r_0}^{r_2} p_f(r',E) dr'\right\}\;,
\end{equation}
where $p_{i,f}$ are the same as in Eq.~(\ref{eq:77}). In fact, the
energy $E$ is equal to the energy of the semiclassical solution,
$E=M$. Indeed, the path integral (\ref{eq:4}) includes integration
over the initial and final configurations of the system, i.e.\ over
$r_1$ and $r_2$ in the shell model. The condition for the stationary
value of $r_1$ reads,
\begin{equation}
\label{eq:34}
\frac{\d}{\d r_1}\log{\Psi_i}+
i\frac{\partial}{\partial r_1}S_{0}(0^-,t_i)  = 0
\qquad \Rightarrow \qquad p_i(r_1,E)=p_i(r_1,M)\;,
\end{equation}
and similarly for $r_2$. This implies 
equality of $E$ and $M$. Note that the parameter $r_0\gg 2M$
in Eq.~(\ref{eq:33}) fixes the phases of the
initial
and final wavefunctions. Namely, the phases vanish at $r=r_0$.

\paragraph{\em The result.}

\begin{figure}[t]
    \centerline{\includegraphics[width=6cm]{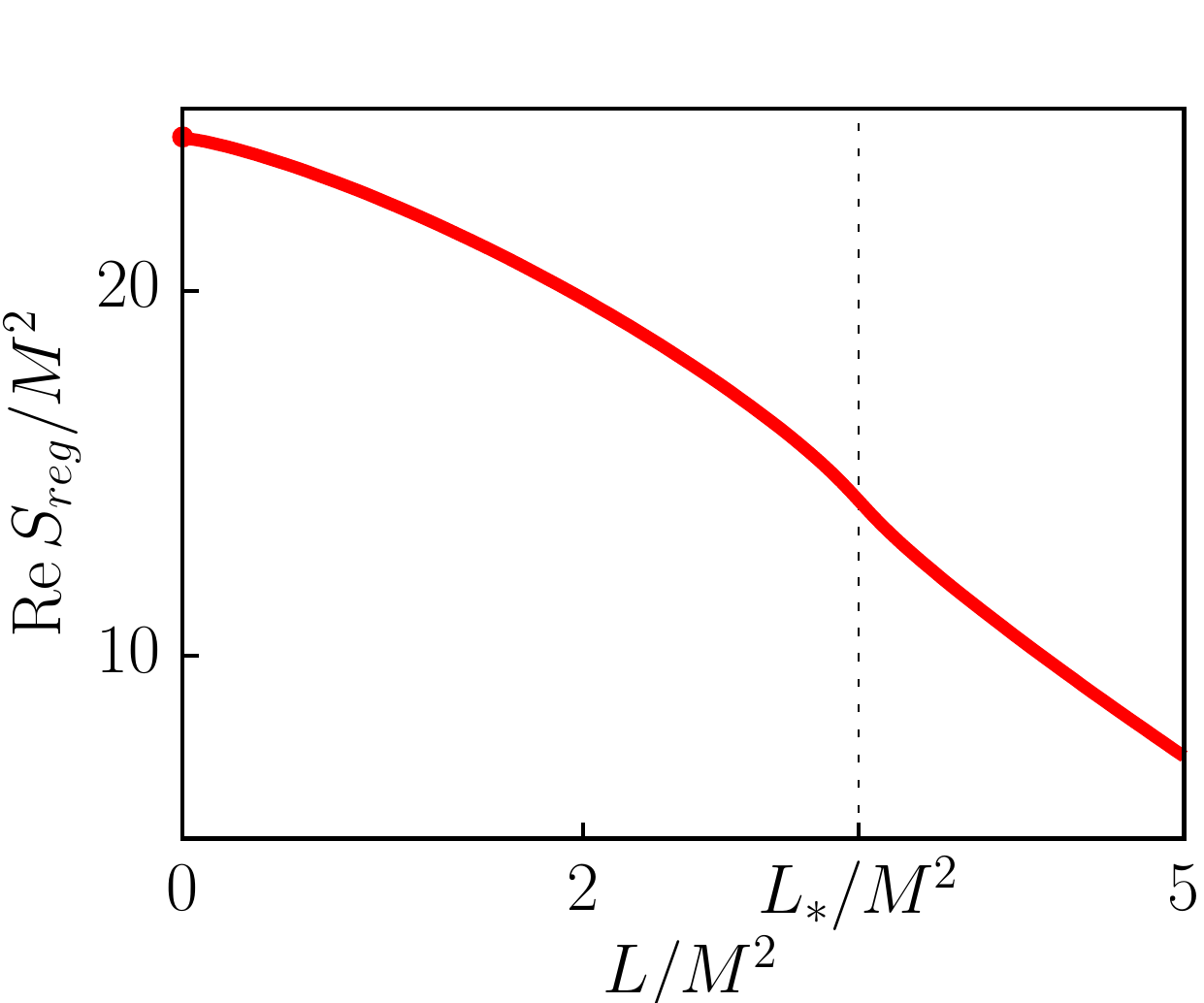}
\hspace{1.8cm}
\includegraphics[width=6cm]{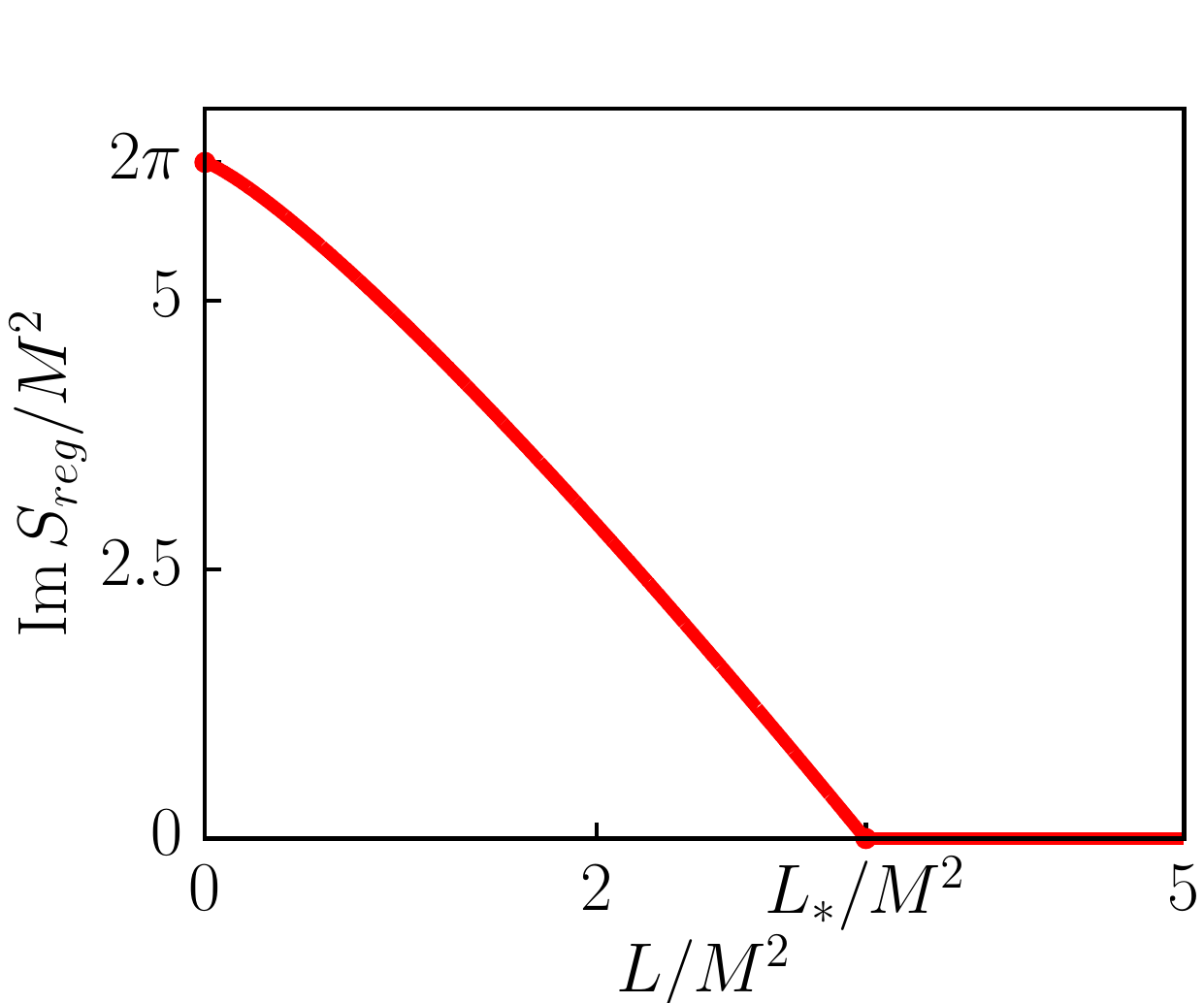}}
\hspace{3.6cm}(a) \hspace{7.4cm} (b)
\caption{Real (a) and imaginary (b) parts of $S_{reg}$ at $m=0$ as
  functions of $L$. For the real part we take $r_0=10 M$.  
\label{fig:ImS_usual}}
\end{figure}

Combining the contributions (\ref{eq:44}), (\ref{eq:33}) with
Eq.~(\ref{eq:Sint}), we obtain the exponent of the ${\cal S}$--matrix
element~\eqref{eq:4}, 
\be
\label{eq:Sreg}
\begin{split}
S_{reg}=&\int\limits_{\cal C}\!\!\frac{dr}{\sqrt{-V_{\rm eff}}}
\bigg[\frac{m^2-2m^2_{\rm eff}}{2 m_{\rm eff}}+
\frac{M}{2}\frac{\sqrt{f_+-V_{\rm eff}}}{f_+}\bigg]\\
&-\sqrt{M^2-m^2}\,\bigg( \frac{r_i+r_f}{2}-2r_0\bigg) 
+\frac{M(2M^2- m^2)}{2\sqrt{M^2-m^2}}\Big(1-\log(r_i r_f/r_0^2)\Big)\,.
\end{split}
\ee
Notice that this result does not depend on any 
regularization parameters.
It is straightforward to check that Eq.~(\ref{eq:Sreg}) is finite in the limit
$r_{i,f}\to+\infty$. In Fig.~\ref{fig:ImS_usual} we plot its 
real and imaginary parts as  functions of $L$ for the case of massless
shell ($m=0$). The imaginary part vanishes for the values $L\geq L_*$
corresponding to classical reflection. At smaller $L$ the imaginary
part is positive implying that the reflection probability
\[
{\cal P}_{fi}\simeq \e^{-2\Im S_{reg}}
\] 
is exponentially suppressed. 
Importantly, $S_{reg}$ does not receive large contributions from the
small--$r$ region near the spacetime singularity. It is therefore  not 
sensitive to the effects of trans--Planckian physics.

In the most interesting case of vanishing centrifugal barrier $L\to
0$ the
only imaginary contribution to $S_{reg}$ comes from the 
residue at the horizon $r_h=2M$ in Eq.~(\ref{eq:Sreg}), recall the contour   
${\cal C}$ in Fig.~\ref{fig:potential}. The respective value of the 
suppression exponent is 
\begin{equation}
\label{eq:48}
2\mathrm{Im}\, S_{reg}\Big|_{L=0} = 2\pi M\, \mathrm{Res}_{\, r= r_h}\,
 \frac{\sqrt{f_+-V_{\rm eff}}}{\sqrt{-V_{\rm eff}}\;f_+} = \pi r_h^2\;.
\end{equation}
This result has important physical implications. First,
Eq.~(\ref{eq:48}) depends only on the total energy $M$ of the shell
but not on its rest mass $m$.  Second, the suppression coincides with the
Bekenstein--Hawking entropy of a black hole with mass $M$. The same
suppression was obtained in \cite{Parikh:1999mf,Berezin:1999nn} for
the probability of emitting the total black hole mass in the form of
a single shell. We conclude that Eq.~(\ref{eq:48})
admits physical interpretation as the probability of the two--stage
reflection process where the black hole is formed in classical
collapse with probability of order 1, and decays afterwards into a
single shell with exponentially suppressed probability. 

One may be puzzled by the fact that, according to Eq.~(\ref{eq:Sreg}), the
suppression receives equal contributions from the two parts of the
shell trajectory crossing the horizon in the inward and outward
directions. Note, however, that the respective parts of the integral
\eqref{eq:Sreg} do not have individual physical meaning. Indeed, we
reduced the original two--dimensional integral for the action to the
form
\eqref{eq:Sreg}
by integrating over sections of constant Schwarzschild time. Another
choice of the sections would lead to an expression with a different
integrand. In particular, using constant--time slices in Painlev\' e
or Finkelstein coordinates one obtains no imaginary contribution to
$S_{reg}$ from the inward motion of the shell, whereas the
contribution from the outward motion is doubled. 
The net result for the probability is, of course, the same.\footnote{Note
 that our semiclassical method is free of uncertainties~\cite{Chowdhury:2006sk,
  Akhmedov:2006pg,  Akhmedov:2008ru} appearing in the approach
of~\cite{Parikh:1999mf}.}

The above result unambiguously shows that the shell model, if taken
seriously as a full quantum theory, suffers from the information
paradox. Indeed, transition between the only two asymptotic states
in this theory --- contracting and expanding shell --- is
exponentially suppressed. Either the theory is intrinsically
non--unitary or one has to take into consideration an
additional asymptotic state of non--evaporating eternal black hole
formed in the scattering process with probability $1-{\cal P}_{fi}$.

On the other hand, the origin of the exponential suppression is clear
if one adopts a modest interpretation of the shell model as describing 
scattering between the narrow wavepackets in field theory. Hawking
effect implies that the black hole decays predominantly into
configurations with high multiplicity of soft quanta. Its decay into a
single hard wavepacket is entropically
suppressed. One can therefore argue~\cite{Parikh:2004ih} 
that the suppression (\ref{eq:48})
is compatible with unitarity of field
theory. However, the analysis of this section is clearly
insufficient to make any conclusive statements in the field theoretic
context. 

As a final remark, let us emphasize that besides the reflection
probability our method allows one to calculate the phase of the
scattering amplitude $\Re S_{reg}$. At $L=m=0$ it can be found
analytically,
\be
\label{eq:ReS}
\Re S_{reg}=2Mr_0+2M^2\log(r_0/2M)+M^2\;.
\ee
Notice that apart from the trivial first term, the phase shift is
  proportional to the entropy $B \propto M^2$; this is compatible with
  the dependence conjectured in~\cite{Giddings:2009gj}.
The phase \eqref{eq:ReS} explicitly depends on the parameter $r_0$ of the initial-- and final--state wavefunctions. 

%%%%%%%%%%%%%%%%%%%%%%%%%%%%%%%%%%%%%%%%%%%%%%%%%%%%%%%%%%%%%%%%%%%%%%%
\subsection{Relation to the Hawking radiation}
\label{sec:relat-hawk-radi-1}
In this section we deviate from the main line of the paper which
studies transitions between free--particle initial and final states, 
and consider scattering of a shell off an eternal
pre--existing black hole. This will allow us to establish a closer
relation of our approach to the results of
\cite{Parikh:1999mf,Berezin:1999nn} and the Hawking radiation. We
will focus on the scattering probability and thus consider only the
imaginary part of the action. 

\begin{figure}[t]
\vspace{-1.5cm}
    \centerline{\includegraphics[width=5cm,angle=45]{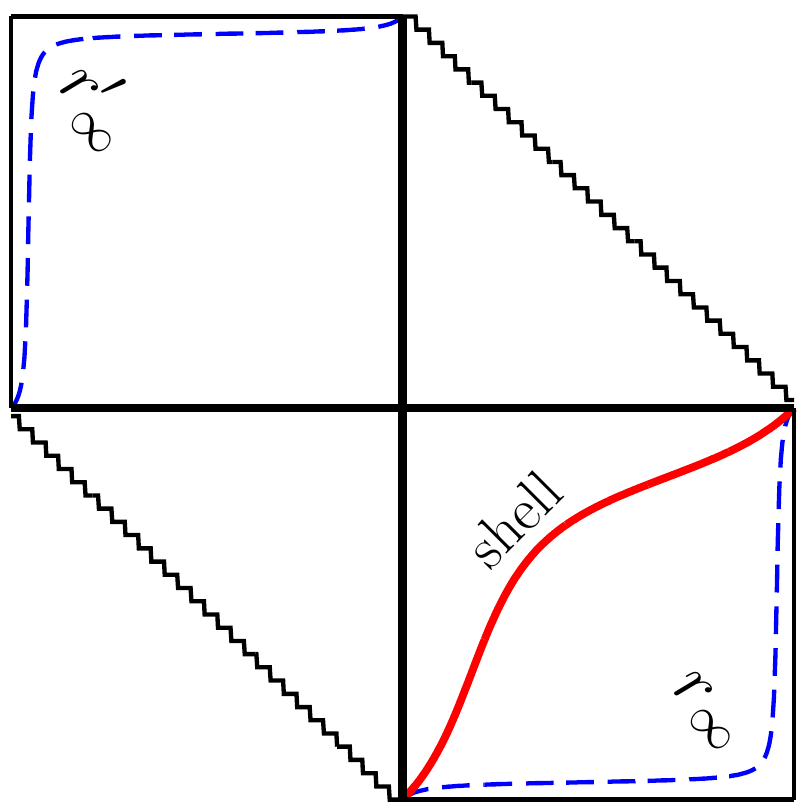}}
\vspace{-1.5cm}
\caption{ Penrose diagram for scattering of a classical shell from an
  eternal black hole. \label{fig:eternalBH}}
\end{figure}

The analysis essentially repeats that of the previous sections with
several differences. First of all, the inner and outer space--times of the
shell are now Schwarzschild with the metric functions
\begin{equation} 
\label{eq:50}
f_{-} = 1 - 2M_{BH}/r \;, \qquad\qquad f_{+} = 1 - 2(M_{BH}+M)/r \;,
\end{equation}
where $M_{BH}$ is the eternal black hole mass and $M$ denotes, as
before, the energy of the shell. The inner and outer metrics possess
horizons at $r_h^{-} = 2M_{BH}$ and $r_h^{+} = 2(M_{BH}+M)$,
respectively. The shell motion is still described by
Eq.~(\ref{eq:30}), where the effective potential is obtained by 
substituting expressions (\ref{eq:50}) into the first line of
Eq.~(\ref{eq:38}). Next,  
the global space--time has an additional boundary $r =
r_{\infty}' \to +\infty$ at the second spatial infinity of the eternal
black hole, see Fig.~\ref{fig:eternalBH}. We have to include 
the corresponding Gibbons--Hawking term, cf. Eq.~(\ref{eq:45}),
\begin{equation}
\label{eq:52}
S_{GH}\Big|_{r_{\infty}'} = -\frac{M_{BH}}{2} \int dt_{-}\;.
\end{equation}
Finally,  the
eternal black hole in the initial and final states contributes into
the free action $S_0$. We use
the Hamiltonian action of an isolated featureless black hole in empty
space--time~\cite{Kuchar:1994zk},
\begin{equation}
\label{eq:53}
S_{0,BH} = -M_{BH}\int dt_{+}\;,
\end{equation}
where, as usual, the time variable coincides with the asymptotic
time\footnote{To compute the phase of the scattering amplitude, one
  should also take into account long--range interaction of the black
  hole with the shell in the initial and final states.}.
 Since we do not equip the black hole with any degrees
of freedom\footnote{The choice of a model for an isolated black
  hole is the main source of uncertainties in scattering
  problems with eternal black holes.}, its
initial-- and final--state wavefunctions are $\Psi_{BH} = 1$.

Adding new terms (\ref{eq:52}), (\ref{eq:53}) to the action
(\ref{eq:Sreg}), one obtains\footnote{Recall
  that the free action enters this formula with the negative sign,
  $S_{0,BH}=M_{BH}(t_f-t_i)$. 
 Note also that the original Gibbons--Hawking term
  (\ref{eq:45}) is proportional to the total ADM mass $M_{BH}+M$.},
\begin{equation}
\label{eq:54}
\mathrm{Im}\, S_{reg} = \mathrm{Im}\, \int\limits_{\cal C}
\frac{dr}{\sqrt{-V_{\rm eff}}} \left[
\frac{m^2 - 2m_{\mathrm{eff}}^2}{2m_{\mathrm{eff}}} 
+ \frac{M_{BH}+M}{2} \, \frac{\sqrt{f_+-V_{\rm eff}}}{f_+} \right.\\\left.-
\frac{M_{BH}}{2} \, \frac{\sqrt{f_--V_{\rm eff}}}{f_-} \right],
\end{equation}
where the integration contour ${\cal C}$ is similar to that in
Fig.~\ref{fig:potential} (lower panel), it bypasses the two
horizons $r^{-}_h$ and $r_{h}^{+}$ in the lower half of complex 
$r$--plane. In the interesting limit of vanishing centrifugal barrier
$L\to 0$ the imaginary part of the action is again given by the
residues at the horizons, 
\begin{align}
\label{eq:57}
2\mathrm{Im}\, S_{reg}\Big|_{L=0} & = 2\pi (M_{BH}+M)\,
\mathrm{Res}_{r= r_h^{+}}\,
 \frac{\sqrt{f_+-V_{\rm eff}}}{\sqrt{-V_{\rm eff}}\,f_+} 
- 2\pi M_{BH}\, \mathrm{Res}_{r= r_h^{-}}\, \frac{\sqrt{f_--V_{\rm
      eff}}}{\sqrt{-V_{\rm eff}}\,f_-}  \notag
  \\ & = \pi \left[(r_h^+)^2 - (r^-_h)^2\right]\;.
\end{align}
Interpretation of this result is similar to
that in the previous section. At the first stage of transition the
black hole swallows the shell with probability of order one and grows
to the mass $M_{BH} + M$. Subsequent emission of the shell with mass $M$
involves suppression
\begin{equation}
\label{eq:51}
{\cal P}_{fi} \simeq \mathrm{e}^{-B_+ + B_-}\;,
\end{equation}
where $B_{\pm}=\pi (r_h^\pm)^2$ are the entropies of the intermediate and
final black holes. This suppression coincides with the results of~\cite{Parikh:1999mf,Berezin:1999nn}.

At $M_{BH}=0$ the process of this section reduces to reflection of a
single self-gravitating shell and expression \eqref{eq:57} coincides
with Eq.~(\ref{eq:48}). In the other limiting case ${M \ll M_{BH}}$ the
shell moves in external black hole metric without back--reaction.
Reflection probability in this case reduces to the Boltzmann exponent
\begin{equation}
\notag
{\cal P}_{fi} \simeq \mathrm{e}^{-M/T_H}\;,
\end{equation}
where we introduced the Hawking temperature $T_H = 1/(8\pi M_{BH})$. One
concludes  that reflection of low--energy shells proceeds via
infall into the black hole 
and Hawking evaporation, whereas at larger $M$ the
probability (\ref{eq:51}) includes back--reaction effects.

%%%%%%%%%%%%%%%%%%%%%%%%%%%%%%%%%%%%%%%%%%%%%%%%%%%%%%%%%%%%%%%%%%%%%%%
\subsection{Space--time picture}
\label{sec:spacetime-picture-1}
\begin{figure}
\vspace{-2.5cm}
\centerline{\includegraphics[width=6cm,angle=45]{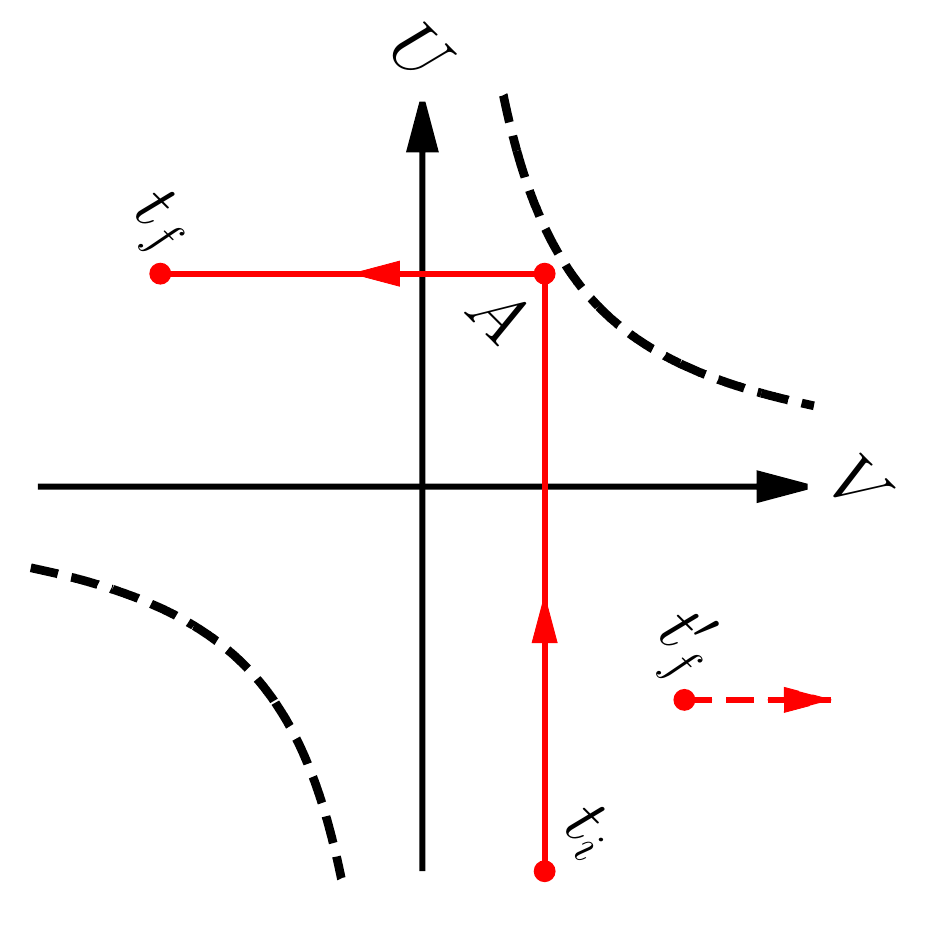}}
\vspace{-2cm}
\caption{Trajectory of the shell in Kruskal coordinates of the outer
  metric. Black dashed lines show the singularities
  $UV=1$.  \label{fig:penrose2}} 
\end{figure}

Let us return to the model with a single shell considered in
Secs.~\ref{sec:models}--\ref{sec:cal-s-matrix}. 
In the previous analysis we integrated
out the non--dynamical metric degrees of freedom and worked 
with the semiclassical shell trajectory $(t_+(\tau),\,
r(\tau))$. It is instructive to visualize this
trajectory in regular coordinates of the outer 
space--time. Below we consider the case of
ultrarelativistic shell with small angular momentum: $L\to 0$
and $M\gg m$. One introduces Kruskal coordinates for the outer
metric,
\begin{equation}
\label{eq:80}
U = - (r/2M' - 1)^{1/2}\, \mathrm{e}^{(r-t_{+})/4M'}\;, \qquad 
V =  (r/2M' - 1)^{1/2}\, \mathrm{e}^{(r+t_{+})/4M'}\;.
\end{equation}
We choose the branch of the square root in these expressions by recalling
that $M'$ differs from the physical energy $M$ by an infinitesimal
imaginary shift, see Eq.~(\ref{eq:35}).  
The initial part of the shell trajectory from $t_+=t_i$ to the turning
point $A$ (Figs.~\ref{fig:potential}, \ref{fig:contour2}) is
approximately mapped to a light ray $V=V_0>0$ as
shown in Fig.~\ref{fig:penrose2}. Note that in the limit $L\to
0$ the turning point $A$ is close to the singularity $r=0$, but does not
coincide with it. At the turning point the shell reflects 
and its radius $r(\tau)$  starts increasing with the proper time
$\tau$. This means that the shell now moves along the light ray
$U=U_0>0$, and the direction of $\tau$ is opposite to that of 
the Kruskal time $U+V$. The corresponding evolution is represented by the
interval $(A,t_f)$ in Fig.~\ref{fig:penrose2}. We conclude that at 
$t_+ = t_f$ the shell emerges in the opposite asymptotic region in the
Kruskal extension of the black hole geometry. This conclusion may seem
puzzling. However, the puzzle is resolved by the 
observation that the two asymptotic regions are related by analytic
continuation in time. Indeed it is  clear from Eqs.~(\ref{eq:80}) that
the shift $t_+\mapsto t_+-4\pi M i$ corresponds to total reflection 
of Kruskal coordinates $U\to -U$, $V\to -V$. Precisely this time--shift
appears if we extend the evolution of the shell to the real time
axis (point $t_f'$ in Fig.~\ref{fig:contour2}). At $t_+ = t_f'$ the 
shell emerges in the right asymptotic region\footnote{Although the
  shell coordinate $r(t_f')$ is complex, we identify the asymptotic
  region using  $\Re r(t_f')$ which is much larger than 
the imaginary part.} 
with future--directed proper
time $\tau$.  The process in Fig.~\ref{fig:penrose2} can be viewed as
a shell--antishell annihilation which is turned by the analytic
continuation into the transition of a single shell from $t_i$ to
$t_f'$. 

Now, we write down the space--time metric for the saddle--point solution 
at $m=0$ and $L\to 0$. Recall that in this case the shell moves along
the real $r$--axis. We therefore introduce 
global complex coordinates $(r,\, t_+)$, where $t_+$ belongs to ${\cal
  C}_t$ and $r$ is real positive. The metric is given by analytic
continuation of Eqs.~(\ref{eq:21}),
(\ref{eq:19}),
\begin{equation}
\label{eq:spmetric}
ds^2=\begin{cases}
-\Big(1-\frac{2M'}{r_{shell}(t_+)}\Big)^2\,dt_+^2+dr^2+r^2d\Omega^2\;,&
r<r_{shell}(t_+)
\\
-\Big(1-\frac{2M'}{r}\Big)\,dt_+^2+\frac{dr^2}{1-\frac{2M'}{r}}+r^2d\Omega^2
\;, & r>r_{shell}(t_+)\;,
\end{cases}
\end{equation}
where we changed the inner time $t_-$ to $t_+$ by matching them at the
shell worldsheet $r = r_{shell}(t_+)$. Importantly, the 
metric  (\ref{eq:spmetric}) is regular at the origin $r=0$ which is
never reached by the shell. It is also well defined at $r_h=2M$ due to the
imaginary part of $M'$; in the vicinity of the Schwarzschild horizon
$r_h$ the metric components are essentially complex. Discontinuity of
Eq.~(\ref{eq:spmetric}) at $r=r_{shell}(t_+)$ is a 
consequence of the $\delta$--function singularity in the shell
energy--momentum tensor. This makes the analytic continuation of the
metric ill--defined in the vicinity of the shell trajectory.
We
expect that this drawback disappears in
the realistic field--theory setup where the
saddle--point metric will be smooth (and complex--valued) 
in Schwarzschild  coordinates. 

%%%%%%%%%%%%%%%%%%%%%%%%%%%%%%%%%%%%%%%%%%%%%%%%%%%%%%%%%%%%%%%%%%%%%%%%
\section{Massless shell in AdS}
\label{sec:massless-shell-ads_4}
\subsection{Reflection probability}
\label{sec:AdS-prob}
In this and subsequent sections we subject our method to further tests
in more complicated shell models. Here we consider a massless shell in
4-dimensional AdS space-time. The analysis is similar to that of
Sec.~\ref{sec:shell-reflections}, so we will go fast over details.

The shell action is still given by Eq.~(\ref{eq:18}) with $m_{\rm
  eff}=L/r$, while the Einstein--Hilbert action is supplemented by
the cosmological constant term, 
\begin{equation}
\label{eq:59}
S_{EH} = \frac{1}{16 \pi} \int_{\cal V} d^4 x\sqrt{-g}\,
(R-2\Lambda)\;.
\end{equation}
Here $\Lambda \equiv -3/l^2$, $l$ is the
AdS radius. The Gibbons--Hawking term has the form (\ref{eq:14}),
where now the regulator at the distant sphere
\begin{equation}
\label{eq:K0AdS}
K_0\big|_{r_\infty} = \frac{2}{l}+\frac{l}{r_{\infty}^2}
\end{equation}
is chosen to cancel the gravitational action of an empty AdS$_4$. The
metric inside and outside the shell is AdS and
AdS--Schwarzschild, respectively,
\begin{equation}
\label{eq:60}
f_{-} =  1 + \frac{r^2}{l^2}\;, \qquad \qquad f_{+} = 1 -
\frac{2M}{r} + \frac{r^2}{l^2}\;, 
\end{equation}
where $M$ is the shell energy. 
The trajectory of the shell obeys Eq.~(\ref{eq:30}) with the effective
potential given by the first line of Eq.~(\ref{eq:38}),
\begin{equation}
\label{eq:VeffAdS}
V_{\rm eff}=1+\frac{r^2}{l^2}-\frac{(L^2+2Mr^3)^2}{4r^4L^2}\;.
\end{equation}
The $\epsilon$--modification again promotes $M$ in this expression to 
$M'=M+i\tilde{\epsilon}$. Repeating the procedure of 
Sec.~\ref{sec:modification}, we start from the reflected trajectory
at large $L$. Keeping  
$\tilde{\epsilon}>0$,  we trace the motion of the turning
point as $L$ decreases\footnote{Alternatively, one can start from the
  flat--space trajectory and continuously deform it by introducing the
  AdS radius $l$.}. The result is a family of contours ${\cal
  C}$ spanned by the trajectory in the complex $r$--plane. These are
similar to the contours in Fig.~\ref{fig:potential}. In particular, at
$L\to 0$ the contour ${\cal C}$ mostly runs along the real axis
encircling the AdS--Schwarzschild horizon $r_h$ from below, as in
the lower panel of Fig.~\ref{fig:potential}.

Calculation of the action is somewhat different from that in flat
space. First, the space--time curvature is now non-zero
everywhere. Trace of the Einstein's equations gives\footnote{In
  the massless case the trace of the shell energy--momentum tensor
  vanishes, $T_{shell\, \mu}^{\phantom{shell}\mu}= 0$, see
  Eqs.~(\ref{eqs:23*}).} $R=4\Lambda$. The Einstein--Hilbert
action takes the form,
\begin{equation}
\label{eq:61}
S_{EH} =\frac{\Lambda}{2}\bigg[\int dt_- \int\limits_0^{r_{shell}}r^2 dr
+ \int dt_+ \int\limits_{r_{shell}}^{r_\infty}r^2 dr\bigg]
=
\int\limits_{shell} \frac{r^3}{2l^2} \left(dt_{+} -
dt_{-}\right) - \frac{r_{\infty}^3}{2l^2} \int dt_{+}\;.
\end{equation}
The last term diverging at $r_\infty\to\infty$ is canceled
by the similar contribution in the Gibbons--Hawking term at spatial
infinity,
\begin{equation}
\label{eq:62}
S_{GH}\big|_{r_\infty} =
\bigg(\frac{r_\infty^3}{2l^2}-\frac{M}{2}\bigg)
\int dt_+\;.
\end{equation}
Second, unlike the case of asymptotically flat space--time, 
Gibbons--Hawking terms at the initial-- and final--time hypersurfaces
$t_+=t_{i,f}$ vanish, see  
Appendix~\ref{sec:gibb-hawk-terms}. Finally, the canonical 
momenta\footnote{They follow from the
   dispersion relation $m_{\rm eff}^2 = -g^{00}M^2-g^{rr}p^2$.} of
the free shell in AdS,
\begin{equation}
p_{i,f}^2(r,M) =  \frac{M^2}{(1+r^2/l^2)^2} - \frac{L^2}{r^2(1+r^2/l^2)}\;,
\end{equation}
are negligible in the asymptotic region $r\to +\infty$. 
Thus, the terms involving $p_{i,f}$ in the free action (\ref{eq:44})
and in the initial and final wavefunctions (\ref{eq:33}) are
vanishingly small if the normalization point $r_0$ is large
enough. This leaves only 
the temporal contributions in the free actions,
\be
\label{eq:S0AdS}
S_0(0^-,t_i)+S_0(t_f,0^+)=M(t_f-t_i)\;.
\ee
Summing up Eqs.~(\ref{eq:61}), (\ref{eq:62}), (\ref{eq:S0AdS}) and the
shell action (\ref{eq:18}), we obtain,
\begin{align}
S_{reg} &= \int_{\cal C} \frac{dr}{\sqrt{-V_{\rm eff}}} 
\bigg[ \frac{r^3}{2l^2} \bigg(
\frac{\sqrt{f_+-V_{\rm eff}}}{f_+} - 
\frac{\sqrt{f_--V_{\rm eff}}}{f_-} \bigg)
+\frac{M}{2}\frac{\sqrt{f_+-V_{\rm eff}}}{f_+}
 -  \frac{L}{r} 
\bigg]\notag\\[1ex]
&\longrightarrow
2\int_0^\infty
dr\bigg[\frac{r^3}{2l^2}\bigg(\frac{1}{f_+}-\frac{1}{f_-}\bigg)+
\frac{M}{2f_+}\bigg]\;\qquad \mbox{at}\qquad L\to 0\;,
\label{eq:63}
\end{align}
where the integration contour in the last expression goes below the
pole at $r=r_h$. The integral (\ref{eq:63}) converges
at infinity due to fast growth of functions $f_+$ and $f_-$. In
particular, this convergence implies that there are no 
gravitational self--interactions of the shell in the initial and final
states due to screening of infrared effects in AdS.

The imaginary part of Eq.~(\ref{eq:63}) gives the exponent of the reflection 
probability. It is related to the residue of the
integrand at $r_h$,
\begin{equation}
2\mathrm{Im}\, S_{reg} =  2\pi  \bigg(\frac{r_h^3}{l^2}  +M\bigg)\,
\mathrm{Res}_{r=r_h}\; \frac{1}{f_+(r)} = \pi r_h^2\;. 
\end{equation}
We again find that the probability is exponentially
suppressed by the black hole entropy. Remarkably, the dependence of the
reflection probability on the model parameters has combined into
$r_h$ which is a complicated function of the AdS--Schwarzschild
parameters $M$ and $l$.

%%%%%%%%%%%%%%%%%%%%%%%%%%%%%%%%%%%%%%%%%%%%%%%%%%%%%%%%%%%%%%%%%%%%%%%
\subsection{AdS/CFT interpretation}
\label{sec:phys-interpr-shells}
Exponential suppression of the shell reflection has a natural
interpretation within the AdS/CFT correspondence~\cite{Maldacena:1997re,
  Witten:1998qj, Gubser:1998bc}. The latter establishes relationship
between gravity in AdS and strongly interacting conformal field theory
(CFT). Consider three--dimensional CFT on a manifold with topology
$\mathbb{R}\times S^2$ parameterized by time $t$ and spherical angles
$\theta$. This is the topology of the AdS$_4$ boundary, so one can think
of the CFT$_3$ as living on this boundary. Let us build the CFT dual
for transitions of a gravitating shell in AdS$_4$. Assume the CFT$_3$
has a marginal scalar operator
$\hat{O}(t,\theta)$; its conformal dimension is $\Delta=3$. This
operator is dual to a massless scalar field $\phi$ in AdS$_4$. 

\begin{figure}[tb]
\centerline{\includegraphics[width=3.8cm]{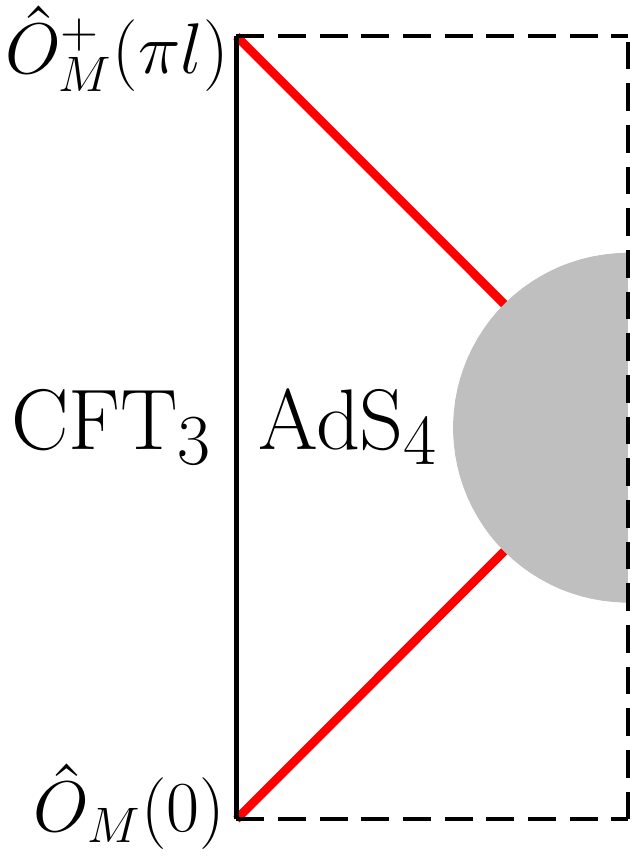}}

\vspace{-2mm}
\caption{Conformal diagram
for scattering of a massless shell in AdS$_4$.
Creation and annihilation of the shell at the AdS boundary correspond to
insertions of composite operators in the CFT dual. \label{fig:AdS}}
\end{figure}

Consider now the composite operator
\be
\label{eq:compop}
\hat O_M(t_0)=\exp \left\{\int dt \, d^2\theta\, G_M(t-t_0)\, \hat{O}(t,\theta)\right\}\;,
\ee
where $G_M(t)$ is a top--hat function of width $\Delta t \gg
1/M$. This operator is dual to a spherical wavepacket (coherent state) of the
$\phi$--field  emitted at time $t_0$ from the boundary towards the
center of AdS~\cite{Balasubramanian:1998de,Polchinski:1999ry}. With an
appropriate 
normalization of $G_M(t)$, the energy of the wavepacket is
 $M$. Similarly, the operator $\hat O_M^+(t_0)$ is dual to
the wavepacket absorbed on the boundary at time $t_0$. Then the
correlator\footnote{The time $\pi l$ is needed for the wavepacket to
  reach the center of AdS and come back if it moves 
classically~\cite{Polchinski:1999ry}.}    
\begin{equation}
\label{eq:56}
{\cal G}_M = \langle \hat O_M^+(\pi l) \,\hat O_M(0)\rangle\;
\end{equation}
is proportional to the amplitude for reflection of the contracting
wavepacket back to the boundary. 
If the width of the wavepacket is small enough, $\Delta t \ll l$, the
$\phi$--field can be treated in 
the eikonal approximation and the wavepacket follows a sharply defined
trajectory. In this way we arrive to the
transition of a massless spherical shell in AdS$_4$, see Fig.~\ref{fig:AdS}.

Exponential suppression of the transition probability implies respective 
suppression of the correlator (\ref{eq:56}). However, the
latter suppression is natural in CFT$_3$ because the state created by the
composite operator $\hat O_M(0)$ is very special. Submitted to time
evolution, it evolves into a thermal equilibrium which poorly 
correlates with the state destroyed by $\hat O^+_M(\pi l)$. 
Restriction of the full quantum theory in AdS$_4$ to  a
single shell is equivalent to a
brute--force amputation of states with many soft quanta in unitary
CFT$_3$. Since the latter are mainly produced during 
thermalization, the amputation procedure leaves us with exponentially
suppressed ${\cal S}$--matrix elements.

%%%%%%%%%%%%%%%%%%%%%%%%%%%%%%%%%%%%%%%%%%%%%%%%%%%%%%%%%%%%%%%%%%%%%%%
\section{Charged shells}
\label{sec:charged-shells}

%%%%%%%%%%%%%%%%%%%%%%%%%%%%%%%%%%%%%%%%%%%%%%%%%%%%%%%%%%%%%%%%%%%%%%%
\subsection{Elementary shell}
\label{sec:prob-trans-1}
Another interesting extension of the shell model is obtained by
endowing the shell with electric charge. The corresponding action is
the sum of Eq.~(\ref{eq:31}) and the electromagnetic contribution
\begin{equation}
\label{eq:82}
S_{EM} = -\frac{1}{16\pi} \int d^4 x \sqrt{-g} \, F_{\mu\nu}^2 - Q
\int_{shell} A_a dy^a\;,
\end{equation}
where $A_\mu$ is the electromagnetic field with stress tensor $F_{\mu\nu}
= \partial_\mu A_{\nu} - \partial_\nu A_\mu$ and $Q$ is the shell
charge. 
This leads to
Reissner--Nordstr\"om (RN) metric outside the shell and empty flat
space--time inside, 
\begin{equation}
\label{eq:70}
f_{+} = 1 - \frac{2M}{r} + \frac{Q^2}{r^2}\;,\qquad A_{0\,+} =
\frac{Q}{r} \;;\qquad\qquad  f_{-} = 1\;, \qquad A_{0\,-} = 0\;.
\end{equation}
Other components of $A_\mu$ are zero everywhere. 
Importantly, the outside metric has two horizons
\begin{equation}
\label{eq:81}
r_{h}^{(\pm)} = M \pm \sqrt{M^2 - Q^2}
\end{equation} 
at $Q<M$. At $Q>M$ the horizons lie in the complex plane, and the
shell reflects classically. Since the latter classical reflections 
proceed without any centrifugal barrier, we
set $L=0$ henceforth. The semiclassical trajectories will be obtained by
continuous change of the shell charge $Q$.

The evolution of the shell is still described by Eq.~(\ref{eq:30}) with the
effective potential constructed from the metric functions
(\ref{eq:70}),
\be
\label{eq:VeffQ}
V_{\rm eff}=1-\frac{(m^2-Q^2+2Mr)^2}{4m^2r^2}\;.
\ee 
This potential always has two turning points on the real axis,
\be
\label{eq:tpQ}
r_{A,A'}=\frac{Q^2-m^2}{2(M\mp m)}\;.
\ee
The shell reflects classically from the rightmost turning
point $r_A$ at $Q>M$. In the opposite case $Q<M$ the turning points
are covered by the horizons, and the real classical solutions describe
black hole formation.

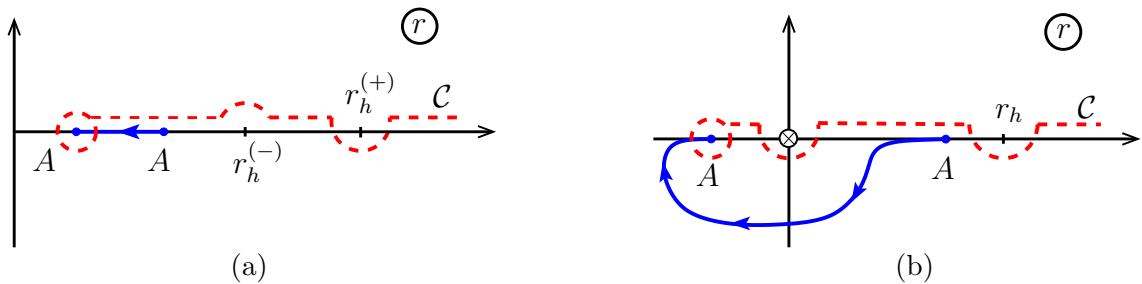
\begin{figure}[t]
\centerline{
\begin{picture}(200,100)
\put(5,0){\includegraphics[width=6.5cm]{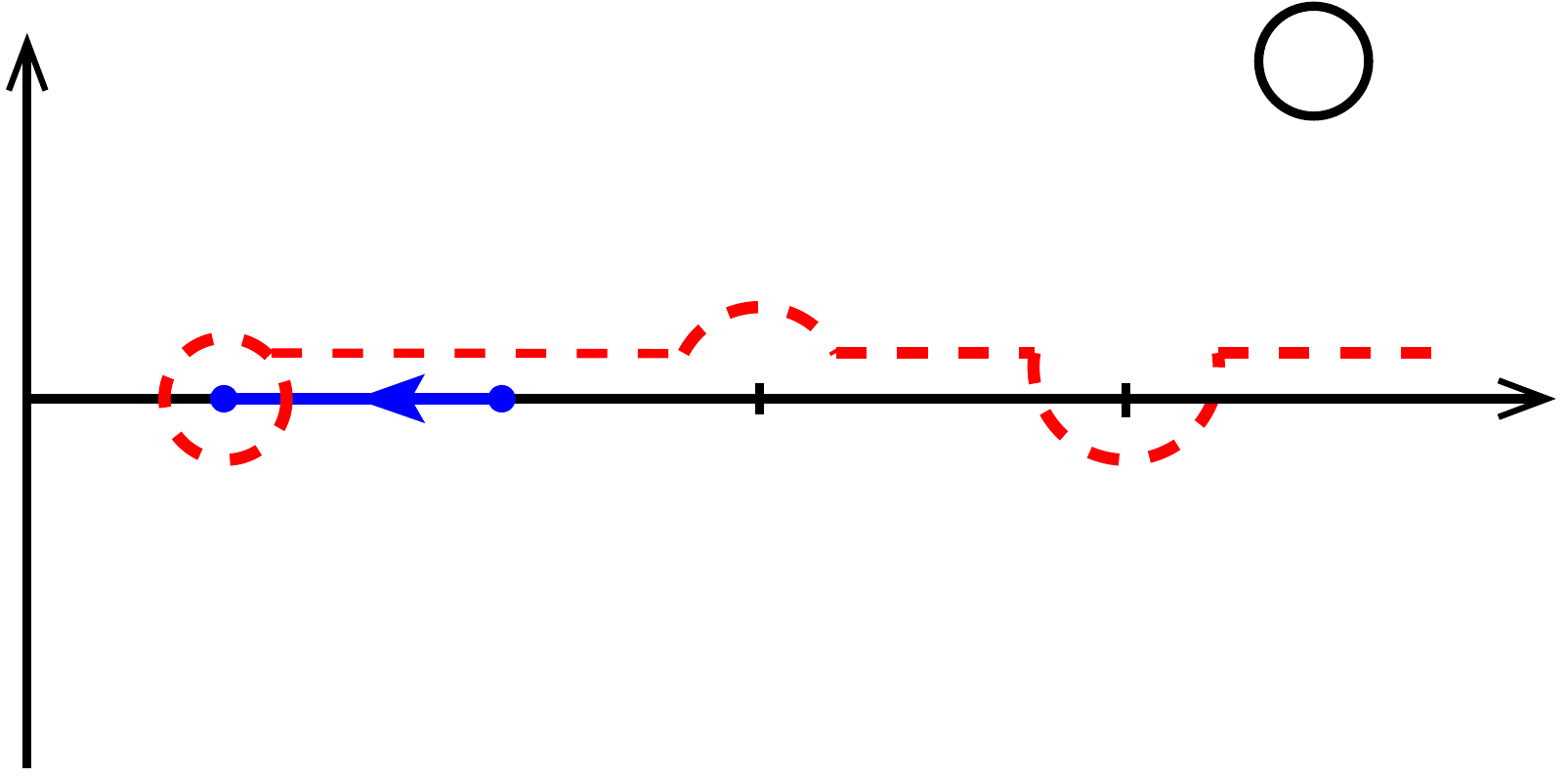}}
\put(157,82){\large $r$}
\put(132,55){\large $r_h^{(+)}$}
\put(90,29){\large $r_h^{(-)}$}
\put(58,30){\large $A$}
\put(15,30){\large $A$}
\put(165,54){\large ${\cal C}$}
\end{picture}
\hspace{1.2cm}
\begin{picture}(200,100)
\put(5,0){\includegraphics[width=6.5cm]{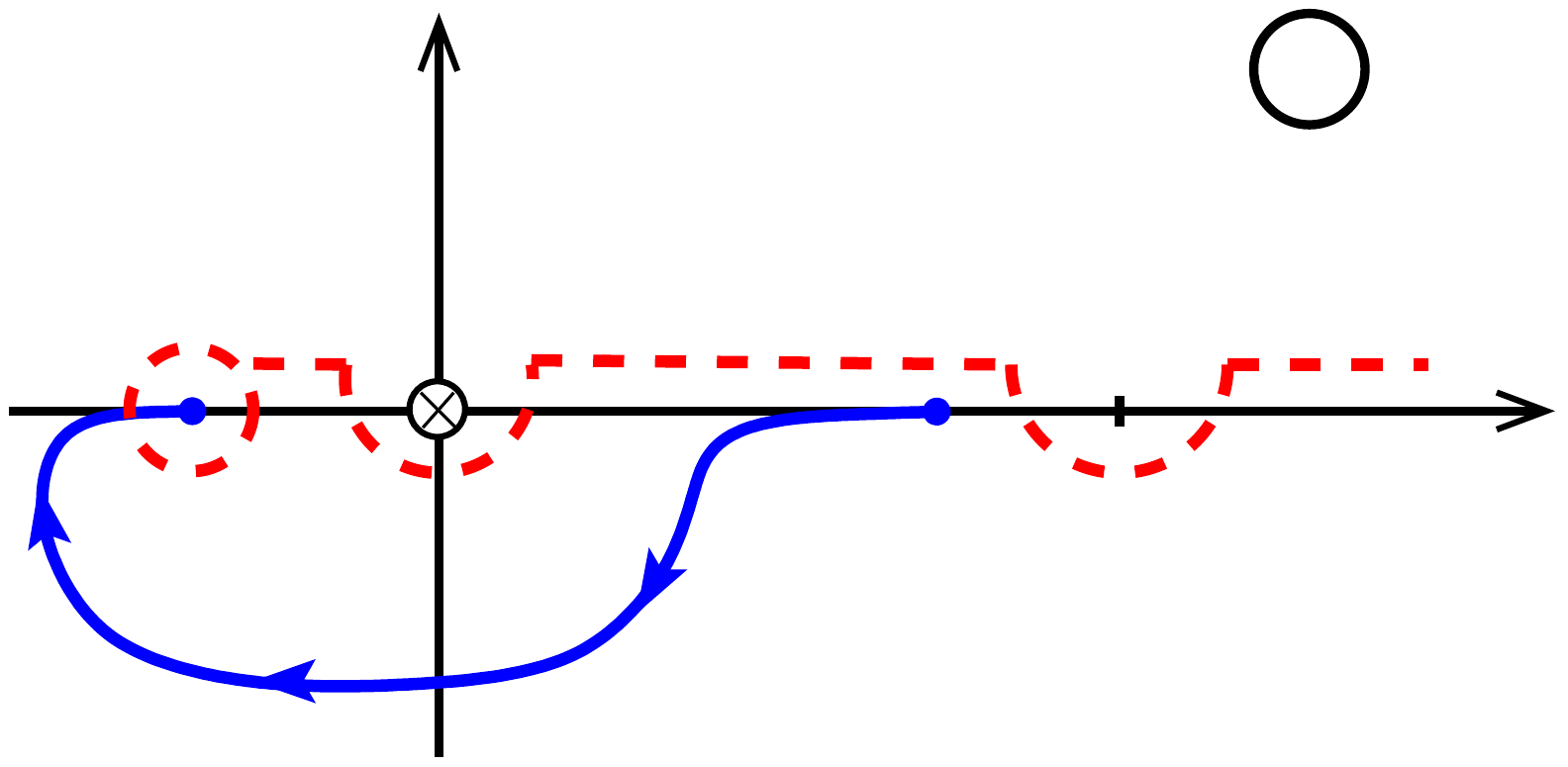}}
\put(157,80){\large $r$}
\put(134,50){\large $r_h$}
\put(22,24){\large $A$}
\put(110,27){\large $A$}
\put(165,50){\large ${\cal C}$}
\end{picture}}
\vspace{-0.2cm}
\hspace{3cm}(a) \hspace{8cm} (b)
\caption{Motion of the turning points and the contour ${\cal C}$ defining
  the trajectory for (a) the model with elementary charged shell and (b)
  the model with discharge.
\label{fig:contour_Q}}
\end{figure}

We find the relevant semiclassical solutions at $Q<M$ using
$\epsilon$--modification. Since the modification term (\ref{eq:10})
does not involve the electromagnetic field, it does not affect the charge
$Q$ giving, as before, an imaginary shift to the mass, $M \mapsto 
M+i\tilde{\epsilon}$. A notable difference from the case of 
Sec.~\ref{sec:shell-reflections} is that the turning points
\eqref{eq:tpQ} are almost real at $Q<M$. The semiclassical
trajectories therefore run close to the real $r$--axis\footnote{The
  overall trajectory is nevertheless complex 
  because $t_+ \in \mathbb{C}$, see below.} for any $Q$. On the other 
hand, the horizons~(\ref{eq:81}) approach the real axis with
$\mathrm{Im}\,  r_h^{(+)} > 0$ and $\mathrm{Im}\, r_h^{(-)}<0$ as $Q$
decreases.  
Thus, the saddle--point trajectories are defined along the contour  ${\cal C}$
in Fig.~\ref{fig:contour_Q}a bypassing 
$r_h^{(+)}$ and  $r_h^{(-)}$ from below and from above,
respectively.

\begin{figure}[t!]
\vspace{-1.5cm}
\begin{center}
\includegraphics[width=5.5cm,angle=45]{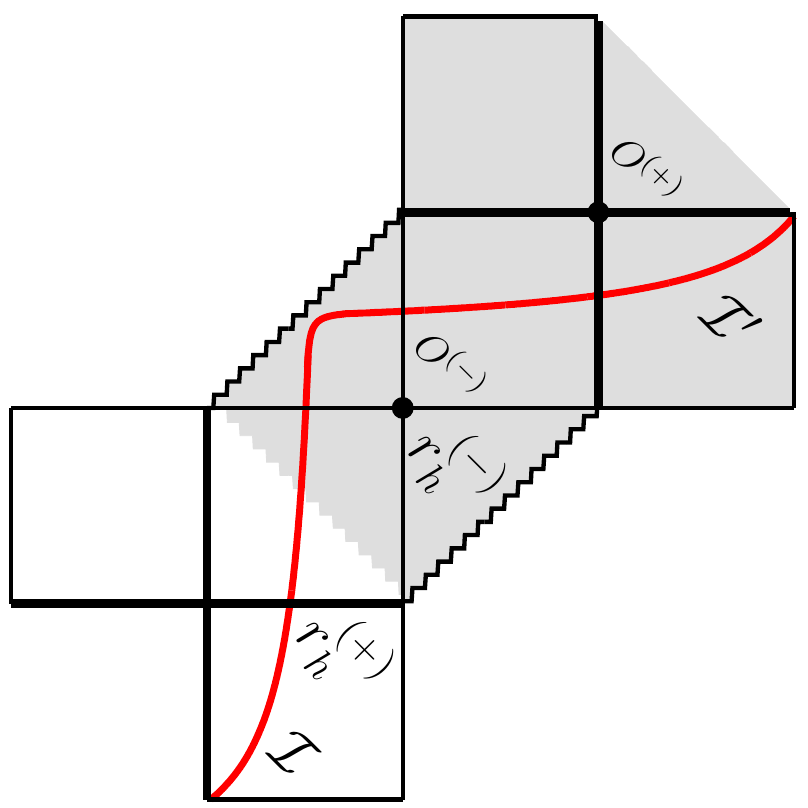}
\end{center}
\vspace{-1cm}
\caption{Conformal diagram for the extended Reissner--Nordstr\"om
  space--time. Semiclassical trajectory running along the contour
  ${\cal C}$ in Fig.~\ref{fig:contour_Q}a is shown by the red line. The grey
  region does not exist in theories with 
  dynamical charged fields.
\label{fig:penrose3}}
\end{figure}

Since the semiclassical motion of the shell at $Q<M$ proceeds with almost real
$r(\tau)$, we can visualize its trajectory in the extended RN geometry, see 
Fig.~\ref{fig:penrose3}. The shell 
starts in the asymptotic region   
${\cal I}$,  crosses the outer and inner horizons $r_h^{(+)}$ and
$r_h^{(-)}$, repels from the time--like singularity due to
electromagnetic interaction, and finally 
re--emerges in the asymptotic region ${\cal I}'$. At first glance, this
trajectory has different topology as compared to the classical reflected
solutions at $Q>M$: the latter stay in the region ${\cal I}$ at the 
final time $t_+= t_f$. However, following
Sec.~\ref{sec:spacetime-picture-1} we recall that 
the Schwarzschild time $t_+$ of the semiclassical trajectory is
complex in the region ${\cal I}'$,
\begin{equation}
\label{eq:17}
\mathrm{Im}\, (t_f-t_i) = \frac{2\pi}{f_+'(r_h^{(+)})} -
\frac{2\pi}{f_+'(r_h^{(-)})}\;, 
\end{equation}
where we used Eq.~\eqref{eq:time} and denoted by $t_i$ and $t_f$ the
values of $t_+$ at the initial and final
endpoints of the contour ${\cal C}$ in
Fig.\ref{fig:contour_Q}a. Continuing $t_f$ to real values, we obtain
the semiclassical trajectory arriving to the region ${\cal I}$ in 
the infinite future\footnote{Indeed, the coordinate systems that are
  regular at the horizons $r_h^{(+)}$ and $r_h^{(-)}$, are
  periodic in the imaginary part of $t_+$ with periods
  $4\pi i/f_+'(r_h^{(+)})$ and $4\pi i/f_+'(r_h^{(-)})$,
  respectively. Analytic continuation to real $t_f$ implies shifts by
  half--periods in both systems, see Eq.~\eqref{eq:17}. This
  corresponds to reflections of the trajectory final point in
  Fig.~\ref{fig:penrose3} with respect to points $O^{(+)}$ and
$O^{(-)}$.}, 
  cf. Sec.~\ref{sec:spacetime-picture-1}. This is what 
one expects  since the asymptotic behavior of the semiclassical
trajectories is not changed in the course of continuous deformations.

Let us now evaluate the reflection probability. Although the contour
${\cal C}$ is real, it receives imaginary contributions from  the
residues at the horizons. Imaginary part of the total action
comes\footnote{Gibbons--Hawking terms at $t =
  t_{i,f}$ are different in the case of charged shell from those in 
  Eq.~(\ref{eq:Sreg}). However, they are real and do not contribute
  into $\mathrm{Im}\, S_{tot}$.} from Eq.~(\ref{eq:Sreg}) and the
electromagnetic term (\ref{eq:82}). The latter takes the form, 
\be
\label{eq:SEM1}
S_{EM}=-\int d^4x\sqrt{-g}\bigg(\frac{F_{\m\n}^2}{16\pi}+A_\m
j^\m\bigg)
=\frac{1}{16\pi}\int d^4x\sqrt{-g}\,F_{\m\n}^2\;,
\ee
where we introduced the shell current $j^\m$, used Maxwell
equations $\nabla_\m F^{\m\n}=4\pi j^\n$ and integrated by parts. 
From Eq.~(\ref{eq:70}) we find,  
\be
\label{eq:SEM2}
S_{EM}=\frac{1}{4}\int dt_+\int\limits_{r_{shell}}^\infty r^2 dr
\bigg(-\frac{2Q^2}{r^4}\bigg) 
=-\frac{Q^2}{2}\int\limits_{shell} \frac{dt_+}{r}\;.
\ee
Combining this with Eq.~(\ref{eq:Sreg}), we obtain,
\begin{align}
2\mathrm{Im}\, S_{reg} =& \mathrm{Im}\,\int_{\cal C}
\frac{dr}{\sqrt{-V_{\rm eff}}} \bigg[ -
m + \bigg(M-\frac{Q^2}{r}\bigg)\frac{\sqrt{f_+-V_{\rm eff}}}{f_+} 
\bigg]\notag\\
\label{eq:83}
=&2\pi\big[\mathrm{Res}_{r=
  r_h^{(+)}}-\mathrm{Res}_{r=
  r_h^{(-)}}\big]
\bigg(M-\frac{Q^2}{r}\bigg)\frac{1}{f_+(r)}
=\pi \big(r_h^{(+)}\big)^2-\pi \big(r_h^{(-)}\big)^2\;.
\end{align}
After non--trivial cancellation we again arrive to a rather simple
expression. However, this time $2\mathrm{Im}\, S_{tot}$ is not equal to
the entropy of the RN black hole, $B_{RN}=\pi 
\big(r_h^{(+)}\big)^2$. 

The physical interpretation of this result is unclear.
We believe that 
it is an artifact of viewing the charged
shell as an elementary object. Indeed, in quantum mechanics of an
  elementary shell the
  reflection probability should vanish at the brink $Q=M$ of classically
allowed transitions. It cannot be equal to $B_{RN}$ which does
not have this property unlike the expression (\ref{eq:83}). 
We now explain how the result is
altered in a more realistic setup.

%%%%%%%%%%%%%%%%%%%%%%%%%%%%%%%%%%%%%%%%%%%%%%%%%%%%%%%%%%%%%%%%%%%%%%%
\subsection{Model with discharge}
\label{sec:models-or-black}
Recall that the inner structure of charged black holes in theories
with dynamical fields is different from the maximal
extension of the RN metric. Namely, the RN Cauchy horizon
$r_h^{(-)}$ suffers from
instability due to mass
inflation and turns into
a singularity~\cite{Poisson:1990eh,Ori:1991zz,Brady:1995ni,Hod:1998gy}. Besides,
pair creation of charged particles forces the singularity to 
discharge~\cite{Novikov:1980ni, Frolov:2005ps,Frolov:2006is}. 
As a result, the geometry near the singularity resembles
that of a Schwarzschild black hole, and the singularity itself is 
space-like. The part of the maximally extended RN space--time including
the Cauchy horizon and beyond (the grey region in
Fig.~\ref{fig:penrose3}) is never formed in classical collapse.

Let us mimic the above discharge phenomenon in the model of a single
shell. Although gauge invariance forbids 
non--conservation of the shell charge $Q$, we can achieve 
essentially the same effect on the space--time geometry by switching
off the electromagnetic interaction at $r\to 0$.
To this end we assume spherical symmetry and introduce a
dependence of the electromagnetic coupling on the radius\footnote{
Alternatively, the discharge can be modeled by introducing 
nonlinear dielectric permittivity~\cite{Sorkin:2000pc}.}. This leads to the
action
\begin{equation}
\label{eq:85}
S_{EM}' = -\frac{1}{16\pi} \int d^4 x \sqrt{-g} \, \frac{F_{\mu\nu}^2}{e^2(r/Q)} \,
 - Q \int\limits_{shell} A_a dy^a\;,
\end{equation}
where $e(x)$ is a positive form--factor starting from 
$e=0$ at $x= 0$ and
approaching $e \to  1$ at $x\to +\infty$. We further assume 
\be
\label{eq:ase}
e(x)<x\;,
\ee
the meaning of this assumption will become clear shortly. Note that
the action (\ref{eq:85}) is invariant  under gauge transformations, as
well as diffeomorphisms preserving the spherical symmetry.
The width of the form--factor $e(r/Q)$ in Eq.~(\ref{eq:85}) scales
linearly with $Q$ to mimic larger discharge regions  
at larger $Q$.

The new action~(\ref{eq:85}) leads to the following solution outside
the shell, 
\begin{equation}
\label{eq:86}
f_{+} = 1 - \frac{2M}{r} + \frac{Q}{r}\, a(r/Q) \;, ~~~
A_{0\, +} = a(r/Q)\;,~~~~ \text{where}~~~~ a(x)
= \int\limits_{x}^{\infty} \frac{e^2(x')}{{x'}^2} \, dx'\;.
\end{equation}
The space--time inside the shell is still empty and flat.  
As expected, the function $f_{+}$ corresponds to the RN metric at large
$r$ and the Schwarzschild one at $r\to 0$. Moreover, the horizon $r_h$
satisfying $f_{+}(r_h) = 0$ is unique due to the condition
(\ref{eq:ase}). 
It starts
from $r_h = 2M$ at $Q=0$,
monotonically decreases with $Q$ and reaches zero at
$Q_{*} = 2M/ a(0)$. At $Q>Q_{*}$ the horizon is absent
and the shell reflects classically.

The subsequent analysis proceeds along the lines
of Secs.~\ref{sec:shell-reflections}, \ref{sec:massless-shell-ads_4}. One
introduces  effective potential for the shell motion,
cf. Eq.~(\ref{eq:VeffQ}), 
\be
\label{eq:Veffdischarge}
V_{\rm eff}=1-\frac{\big(2Mr-Qr\,a(r/Q)+m^2\big)^2}{4m^2r^2}\;,
\ee
introduces $\epsilon$--regularization, $M\mapsto
M'=M+i\tilde\epsilon$, and studies motion of the turning points of
the shell trajectory as $Q$ decreases. If $M,Q\gg m$, this analysis
can be performed for general $e(x)$. In this case the relevant turning
point $r_A$ is real and positive for $Q>Q_*$.  At $Q\approx Q_*$ it comes to
the vicinity of the origin $r=0$ where the function $a(r/Q)$ can be
expanded up to the linear term. The position of the turning
point is
\be
\label{eq:tpdis}
r_A=\frac{1}{b^2}\bigg[-M+m+\frac{a(0)Q}{2}+
\sqrt{\bigg(M-m-\frac{a(0)Q}{2}\bigg)^2-m^2b^2}\bigg]\;,
\ee 
where $b^2 \equiv -da/dx\big|_{x=0}$ is positive according to
Eq.~(\ref{eq:86}). As $Q$ decreases within the interval 
\be
Q_*-\frac{2m(1-b)}{a(0)}>Q>Q_*-\frac{2m(1+b)}{a(0)}\;
\ee
the turning point makes an
excursion into the lower half of the $r$--plane,  
goes below the origin and returns to the real axis on the negative
side, see Fig~\ref{fig:contour_Q}b. For smaller charges $r_A$ is small
and stays on
the negative real axis. The contour
${\cal C}$ defining the trajectory is shown in Fig.~\ref{fig:contour_Q}b. It
bypasses the horizon $r_h$ from below, goes close to the
singularity, encircles the turning point and returns back to
infinity. This behavior is analogous to that in the case of
neutral shell.

Finally, we evaluate the imaginary part of the action. The
electromagnetic contribution is similar to Eq.~(\ref{eq:SEM2}),
\be
\label{eq:SEMprim1}
S_{EM}'=-\frac{Q^2}{2}\int\limits_{shell}\frac{e^2\, dt_+}{r}\;.
\ee 
However, in contrast to Sec.~\ref{sec:prob-trans-1}, the trace of the
gauge field energy--momentum tensor does not vanish due to explicit
dependence of the gauge coupling on $r$ (cf. Eq.~(\ref{eq:23})), 
\be
\label{eq:TEMprim}
{T'^{\phantom{Em}\mu}_{EM\, \mu}}=\frac{F_{\m\n}^2r}{8\pi e^3}\frac{de}{dr}\;.
\ee
This produces non--zero scalar curvature $R=-8\pi
{T'^{\phantom{Em}\mu}_{EM\,\mu}}$ in the outer region of the
shell, and the Einstein--Hilbert
action receives an additional contribution, 
\be
\label{eq:SEHprim}
\Delta S_{EH}=-\frac{1}{4}\int dt_+\int_{r_{shell}}^{\infty}
r^2 dr \bigg(-\frac{2Q^2e^4}{r^4}\bigg)\frac{r}{e^3}\frac{de}{dr}
=\int\limits_{shell} dt_+\bigg(\frac{Q^2e^2}{4r}-\frac{Qa}{4}\bigg)\;,
\ee
where in the second equality we integrated by parts. Combining
everything together, we obtain (cf. Eq.~(\ref{eq:83})),
\begin{align}
2\mathrm{Im}\, S_{reg} &= \mathrm{Im}\,\int_{\cal C}
\frac{dr}{\sqrt{-V_{\rm eff}}} \bigg[-
m+\bigg(M -\frac{Q^2 e^2}{2r} -
\frac{Qa}{2}\bigg) \frac{\sqrt{f_+-V_{\rm eff}}}{f_+}
\bigg]\;,\notag\\
\label{eq:87}
&=2\pi \mathrm {Res}_{r= r_h} \bigg(M - \frac{Q^2
e^2}{2r} - \frac{Qa}{2}\bigg)\frac{1}{f_+} = \pi r_h^2\;,
\end{align}
where non--trivial cancellation happens in the last equality 
for any $e(x)$. To sum up, we accounted for the discharge of the black hole
singularity and recovered the intuitive result: the reflection
probability is suppressed by the entropy of the intermediate black
hole\footnote{We do not discuss the phase of
  the scattering amplitude as it essentially depends on our choice of
  the discharge model.}.

%%%%%%%%%%%%%%%%%%%%%%%%%%%%%%%%%%%%%%%%%%%%%%%%%%%%%%%%%%%%%%%%%%%%%%%
\section{Conclusions and outlook}
\label{sec:concl-prosp}
In this paper we proposed a semiclassical method to calculate the
${\cal S}$--matrix elements for the two--stage transitions involving 
collapse of multiparticle states into a black hole and decay of
the latter into free particles. Our semiclassical
  approach does not require full quantization of
  gravity. Nevertheless, it consistently incorporates backreaction of
  the collapsing and emitted quanta on the geometry.
It reduces evaluation of the ${\cal S}$--matrix elements to finding
complex--valued solutions of the coupled classical Einstein and matter field
  equations with certain boundary conditions.

An important technical ingredient of the method is the 
regularization 
 enforcing the semiclassical solutions to interpolate
  between the in- and out- asymptotic states consisting of free
  particles in flat space--time. 
As a consequence, one works with the complete semiclassical solutions
describing formation and decay of the intermediate black hole.
This distinguishes our approach from the
  standard semiclassical expansion in the black hole
  background. In addition, the same regularization allows us to select
  the relevant semiclassical configurations by continuous deformation of
  the real 
  solutions describing classical scattering at lower energies.
  The final result for the ${\cal S}$--matrix elements
  does not dependent on the details of the regularization.

We illustrated the method in a number of toy models with matter
  in the form of thin shells. We have found that the relevant
  semiclassical solutions are complex--valued and defined in the
  complexified space--time in the case of black hole--mediated
  processes. They avoid the high--curvature region near the black hole
  singularity thus justifying our use of the semiclassical low--energy
  gravity. In particular, the
  Planck--scale physics near the black hole singularity is irrelevant
  for the processes considered in this paper.

The method has yielded sensible results for
  transition amplitudes in the shell models.
Namely, we have found that the probabilities of the two--stage shell transitions
are exponentially suppressed by the Be\-ken\-stein--Hawking entropies of
the  intermediate black holes. If the shell model is taken seriously as
a full quantum theory, this result implies that its 
${\cal
  S}$--matrix is non-unitary. 
However, the same result is natural and consistent with
unitarity if the shells are interpreted as describing 
scatterings of narrow wavepackets  
in field theory. 
The exponential suppression appears because we consider a very
special exclusive process: formation of a black hole by a sharp
wavepacket followed by its decay into the same packet.
Our result  coincides with the probability of black hole
decay into a single shell found within the tunneling approach to 
Hawking radiation \cite{Parikh:1999mf,Berezin:1999nn} and is
consistent with interpretation of the Bekenstein--Hawking entropy
as the number of black hole microstates
\cite{Parikh:2004ih}. Considering the shell in AdS$_4$ space--time we
discussed the result from the AdS/CFT viewpoint. 
 
In the case of charged shells our method reproduces the entropy
suppression only if instability of the Reissner--Nordstr\"om Cauchy
horizon with respect to pair--production of charged particles  is
taken into account. This suggests that the latter process is crucial
for unitarity of transitions with charged black holes at the
intermediate stages. 

Besides the overall probability, our method yields the phase
  of the transition amplitude. The latter carries
  important information about the scattering process, in particular,
  about its initial and final states. In the case of a
  neutral shell in asymptotically flat space--time the
  phase contains a logarithmically divergent term due to
  long--range Newtonian interactions 
  and
 terms proportional 
to the black hole entropy. This is consistent with the behavior
conjectured in \cite{Giddings:2009gj}. 

We consider the above successes as an encouraging confirmation of the
viability of our approach.

The shell models are too simple to address many interesting
  questions about the black hole ${\cal S}$--matrix. These
  include the expected growth of the transition probability when the
  final state approaches the Hawking--like state with many 
quanta, and the recent
  conjecture about sensitivity of the amplitudes to small changes in
  the initial and final states \cite{Polchinski:2015cea}. A study of
  these issues will require application of our method to a genuinely
  field--theoretic setup. 
Let us anticipate the scheme of such analysis. As an example, 
consider a 
 scalar field
$\phi$ minimally coupled to gravity. 
For simplicity, one can restrict to transitions between 
 initial and final
states with particles in the $s$--wave. These states are invariant under rotations.
Then the respective semiclassical solutions are also expected to
possess this symmetry. 
They satisfy the complexified wave-- and Einstein
equations in the spherically symmetric model of gravity 
plus a scalar 
field\footnote{Another interesting arena
  for application of the semiclassical method is two--dimensional dilaton
  gravity~\cite{Callan:1992rs}.}. 
One can use the simplest
Schwarzschild coordinates $(t,\, r)$ which are well--defined for 
complex $r$ and $t$, though other
  coordinate systems may be convenient for practical reasons.
One
starts from wavepackets with small amplitudes $\phi_0$ which scatter
trivially in flat space--time. Then one adds the
complex term (\ref{eq:7}), (\ref{eq:10}) to the classical action  and
finds the modified saddle--point solutions. Finally, one
increases $\phi_0$ and obtains saddle--point solutions for the black
hole--mediated transitions. The space--time manifold, if needed, should be
deformed to complex values of coordinates~--- away from the singularities of
the solutions. 
We argued  in Sec.~\ref{sec:method} that the modified solutions
are guaranteed to approach flat space--time at $t\to +\infty$ and as such, describe
scattering. The ${\cal S}$--matrix element~(\ref{eq:2}) is then
related to the saddle--point action $S_{reg}$ in the limit of
vanishing modification $\epsilon\to +0$. 
Evaluation of ${\cal S}$--matrix elements is thus reduced
to solution of two--dimensional 
complexified field equations, which can be performed on the
present--day computers. 

One may be sceptical about the restriction to the spherically
symmetric sector which leaves out a large portion of the original Hilbert
space. In particular, all states containing
gravitons are dropped off because a massless spin-2 particle cannot be in an 
$s$--wave. Nevertheless, the ${\cal S}$--matrix in this sector is
likely to be rich enough to provide valuable information about the
properties of black hole--mediated scattering. In particular, it is 
sufficient for addressing the questions mentioned in the previous paragraph.

Furthermore, one can envisage tests of the ${\cal S}$--matrix
  unitarity purely within the semiclassical approach.
Indeed, consider the
  matrix element of the operator ${\cal S}^\dagger{\cal S}$ between
  two coherent states with the mode amplitudes $a_k$ and
  $b_k$ ($k$ is the mode wavenumber),
\be
\label{lasteq}
\bra{a}{\cal S}^\dagger{\cal S}\ket{b}=
\int {\cal D}c_k{\cal D}c_k^*\e^{-\int dk\,c_k^*c_k}
\bra{a}{\cal S}^\dagger\ket{c}\bra{c}{\cal S}\ket{b}\;,
\ee
where on the r.h.s.\ we inserted the sum over the (over-)complete set
of intermediate coherent states. For sifficiently distinct
semiclassical states $\ket{a}$ and $\ket{b}$ the integrand in
Eq.~(\ref{lasteq}) is a rapidly oscillating function. Then, it is natural
to assume that the integral will be saturated by a unique saddle--point
state $\ket{c_0}$ which does not coincide with the dominant final
states of the transitions starting from $\ket{a}$ and $\ket{b}$. 
This suggests that the amplitudes $\bra{a}{\cal
  S}^\dagger\ket{c_0}$ and
 $\bra{c_0}{\cal S}\ket{b}$ correspond to rare exclusive processes and 
can be evaluated semiclassically. 
Substituting them in Eq.~(\ref{lasteq}) and comparing 
the
result with the matrix element of the unity operator,
\begin{equation}
\label{eq:78}
\langle a | 1 | b\rangle = \mathrm{e}^{\,\int dk \, a_k^* b_k}\;,
\end{equation}  
one will perform a strong unitarity test for the gravitational ${\cal
  S}$--matrix. 
Of course, this discussion relies on several speculative 
assumptions that must be verified. We plan to return to this
subject in the future.

Finally, in this paper we have focused on the scattering processes
with fixed initial and final states which are described in the {\it
  in-out} formalism. In principle, the semiclassical approach can be
also applied to other quantities, e.g.\ the results of measurements
performed by an observer infalling with the collapsing matter. The
latter quantitites are naturally defined in the {\it in-in} formalism
with the corresponding modification of the path integral. One can
evaluate the new path integral using the saddle--point
technique. Importantly, the new semiclassical solutions need not
coincide with those appearing in the calculation of the
$S$--matrix. If they turn out to be different, it would imply that the
infalling and outside observers describe the collapse/evaporation
process with different semiclassical geometries. This will be an
interesting test of the the black hole complementarity principle. We
leave the study of this exciting topic for future.

%%%%%%%%%%%%%%%%%%%%%%%%%%%%%%%%%%%%%%%%%%%%%%%%%%%%%%%%%%%%%%%%%%%%%%%%
\acknowledgments
We are grateful to D.~Blas, V.~Berezin, R.~Brustein, S.~Dubovsky, G.~Dvali, 
M.~Fitkevich, V.~Frolov, J.~Garriga, 
V.~Mukhanov, V.~Rubakov, A.~Smirnov, I.~Tkachev and T.~Vachaspati for useful
discussions. We thank A.~Barvinsky, A.~Boyarsky and A.~Vikman for
encouraging interest.
This work was supported by the RFBR grant
12-02-01203-a (DL) and the Swiss National
Science Foundation (SS).

%%%%%%%%%%%%%%%%%%%%%%%%%%%%%%%%%%%%%%%%%%%%%%%%%%%%%%%%%%%%%%%%%%%%%%%
\appendix
\section{A shell of rotating dust particles}
\label{sec:shell-rotating-dust}
Consider a collection of dust particles uniformly distributed on a
sphere. Each partice has mass $\delta m$ and absolute value $\delta L$
of angular momentum.
We assume no preferred
direction in particle velocities, so that their angular momenta
sum up to zero. This configuration is
spherically--symmetric, as well as the collective gravitational field.
Since the spherical symmetry is preserved in the course of
classical evolution, the particles remain distributed on the sphere of 
radius $r(\tau)$ at any time $\tau$ forming an infinitely thin shell.

Each particle is described by the action
\begin{equation}
\label{eq:13}
\delta S = -\delta m \int |ds| = - \delta m \int d\tau 
\sqrt{-\mathrm{g}_{ab} \dot{y}^a
  \dot{y}^b - r^2(\tau) \dot{\varphi}^2 }\;,
\end{equation}
where 
in the second equality we substituted the spherically symmetric metric 
(\ref{eq:9})
and introduced the time parameter $\tau$.
To construct the action for $r(\tau)$, we integrate out the
motion of the particle along the angular variable $\varphi$ using 
conservation of angular momentum
\begin{equation}
\label{angmom}
\delta L = \frac{\delta m r^2 
\dot{\varphi}}{\sqrt{-\mathrm{g}_{ab}\dot y^a\dot y^b - r^2
    \dot{\varphi}^2 }}\;.
\end{equation}
It would be incorrect to express $\dot\varphi$ from this formula and
substitute it into Eq.~(\ref{eq:13}). To preserve the equations of
motion, we perform the substitution in the Hamiltonian
\be
\label{Ham}
\delta H=p_a{\dot y}^a +\delta L\dot\varphi-\delta {\cal L}\;,
\ee
where $p_a$ and $\delta L$ are the canonical momenta for 
$y^a$ and $\varphi$, whereas
$\delta {\cal L}$ is the Lagrangian in Eq.~(\ref{eq:13}).
Expressing $\dot\varphi$ from Eq.~(\ref{angmom}), we obtain,
\be
\label{Ham1}
\delta H=p_a{\dot y}^a +\sqrt{-\mathrm{g}_{ab}\dot y^a\dot y^b}
\sqrt{\delta m^2+\delta L^2/r^2}\;.
\ee
From this expression one reads off the action for $r(\tau)$,
\be
\label{eq:15}
\delta \tilde{S} =
- \int d\tau \sqrt{\delta m^2 + \delta L^2/r^2}\;,
\end{equation}
where we fixed $\tau$ to be the proper time along the shell. 
We finally sum up the actions (\ref{eq:15}) of individual particles 
into the shell action
\begin{equation}
\label{eq:16}
S_{shell} = N \delta \tilde{S} = - \int d\tau \sqrt{
  m^2 + L^2/r^2(\tau)}\;,
\end{equation}
where $N$ is the number of particles, $m = N\delta m$ is their total
mass and $L = N\delta L$ is the sum of absolute values of the particles'
angular momenta. We stress that $L$ is {\em not} the total 
angular  momentum of the shell. The latter is zero because the
particles rotate 
in different directions.

%%%%%%%%%%%%%%%%%%%%%%%%%%%%%%%%%%%%%%%%%%%%%%%%%%%%%%%%%%%%%%%%%%%%
\section{Equation of motion for the shell}
\label{sec:israel-condition}
In this appendix we derive equation of motion for the model with the
 action (\ref{eq:31}). We start by obtaining expression for the shell
 energy--momentum tensor. Let us introduce coordinates 
$(y^a,\,\theta^\alpha)$ such that the 
metric~\eqref{eq:9} is continuous\footnote{Schwarzschild coordinates in Eq.~(\ref{eq:21}) 
  are discontinuous at the shell worldsheet.} across the shell. Here
$\theta^\a$, $\a=2,3$ are the spherical angles. Using the identity 
\begin{equation}
\label{ident}
\int d^2y\;\frac{d^2\Omega}{4\pi}\;\delta^{(2)}(y-y(\tau))=1\;,
\end{equation}
we recast the shell action (\ref{eq:18}) as
an integral over the four--dimensional space--time,
\begin{equation}
\label{eq:22}
S_{shell} = -\int d^2y\frac{d^2\Omega}{4\pi}\int d\tau\;
m_\mathrm{eff}\sqrt{-\mathrm{g}_{ab}\dot y^a\dot y^b}\;
\delta^{(2)}(y-y(\tau))\;.
\end{equation}
Here $\tau$ is regarded as a general time parameter. 
The energy--momentum tensor of the shell is obtained by varying
Eq.~(\ref{eq:22})  with respect to $\mathrm{g}_{ab}$ and $r^2(y)$,
\begin{subequations}
\label{eqs:23*}
\begin{align}
\label{eq:230}
&T^{ab}_{shell}=\frac{2}{\sqrt{-g}}\frac{\delta S_{shell}}{\delta
   \mathrm{g}_{ab}}
=\dot y^a\dot y^b \frac{m_{\mathrm{eff}}}{4\pi r^2} 
\int d\tau\;\frac{\delta^{(2)}(y-y(\tau))}{\sqrt{-\mathrm{g}}}\;,
\\
\label{eq:23}
&T^{\phantom{shell}\alpha}_{shell\,\alpha} = \frac{2 r^2}{\sqrt{-g}} \; \frac{\delta
  S_{shell}}{\delta r^2} = \frac{L^2}{4\pi r^4 m_{\mathrm{eff}}}
\int d\tau\;\frac{\delta^{(2)}(y-y(\tau))}{\sqrt{-\mathrm{g}}}
\;,
\end{align}
\end{subequations}
where in the final expressions we again set $\tau$ equal to the proper
time. It is straightforward to see that the $\tau$--integrals in
Eqs.~\eqref{eqs:23*} produce $\delta$--functions of the geodesic
distance $n$ from the shell,
\begin{equation}
\label{deltan}
\delta(n)=\int
d\tau\,\frac{\delta^{(2)}(y-y(\tau))}{\sqrt{-\mathrm{g}}}\;.
\end{equation}
We finally arrive at 
\begin{equation}
\label{eq:20}
T_{shell}^{\m\n}=t^{\m\n}_{shell}\, \delta(n)\;,~~~~~~~~~
t^{ab}_{shell} = \frac{m_{\mathrm{eff}} \dot{y}^a \dot{y}^b}{4\pi
  r^2}~,~~~~~~~
t_{shell\,\beta}^{\phantom{shell}\,\alpha} =
\delta^\alpha_\beta\,\frac{L^2}{8\pi m_{\mathrm{eff}} r^4 }\;,
\end{equation}
where $T_{shell\,\b}^{\phantom{shell}\a}\propto\delta^\a_\b$ due to
spherical symmetry.

Equation of motion for the shell is the consequence of Israel junction
conditions which follow from the Einstein equations.  
The latter conditions relate $t_{shell}^{\m\n}$ to the jump in the
extrinsic curvature across the shell
\cite{Israel:1966rt, Berezin:1987bc} 
\begin{equation}
\label{eq:24}
(K^\mu_\nu)_+-(K^\mu_\nu)_- 
= -8\pi\, \bigg(t_{shell\,\nu}^{\phantom{shell}\,\mu} - \frac12
h^\mu_\nu\, t_{shell\, \lambda}^{\phantom{shell\,}\lambda}\bigg)\;.
\end{equation}
Here $h^\m_\n$ is the induced metric on the shell, $K_{\m\n}$ is its
extrinsic curvature, the subscripts $\pm$ denote quantities 
outside ($+$) and inside ($-$) the shell. 
We define both $(K_{\m\n})_\pm$ 
using the outward--pointing normal, $n_\m \d_r
    x^\m>0$.
Transforming the metric \eqref{eq:21}
into the continuous coordinate system, we obtain,
\begin{equation}
\label{extcurv}
(K^{ab})_{\pm}=-\dot y^a\dot y^b\frac{\ddot r+f'_\pm/2}{\sqrt{\dot
    r^2+f_{\pm}}}~,~~~~~ 
(K^\a_\b)_{\pm}=\delta^\a_\b\frac{\sqrt{\dot r^2+f_\pm}}{r}\;,
\end{equation}
where dot means derivative with respect to $\tau$. 
From Eq.~(\ref{eq:24}) we derive the equations,
\begin{align}
\label{eq:29}
& \sqrt{\dot{r}^2 + f_{+}} - \sqrt{\dot{r}^2 + f_{-}} =
  -\frac{m_{\mathrm{eff}}}{r}\;,\\ 
\label{eq:28}
& \frac{\ddot{r} + f'_{+}/2}{\sqrt{\dot{r}^2 + f_{+}}} -
  \frac{\ddot{r} + f'_{-}/2}{\sqrt{\dot{r}^2 + f_{-}}} =
  \frac{L^2}{m_{\mathrm{eff}}r^4} + \frac{m_{\mathrm{eff}}}{r^2}\;.
\end{align} 
Only the first equation is independent, since the second is
proportional to its time derivative. We conclude that Einstein
equations are fulfilled in the entire space--time provided the metrics
inside and outside the shell are given by Eqs.~(\ref{eq:21}),
(\ref{eq:19}) and Eq.~(\ref{eq:29}) holds at the shell worldsheet.
The latter equation is equivalent to
Eqs.~(\ref{eq:30}), (\ref{eq:38}) from the main text.

The action (\ref{eq:31}) must be also extremized with respect to
the shell trajectory $y^a(\tau)$. However, the resulting
equation is a consequence of Eq.~(\ref{eq:29}). 
Indeed, the shell is described by a single coordinate $r(\tau)$, and
its equations of motion are equivalent to  conservation of the
energy--momentum tensor. The latter conservation, however, is 
ensured by the Einstein equations. 

\section{Turning points at $L\to 0$}
\label{app:tp}
Turning points are zeros of the effective potential
$V_{\mathrm{eff}}$, Eq.~(\ref{eq:38}). The
latter has six zeros $r_i$, ${i = 1\dots 6}$, which can be expressed
analytically at $L\to 0$. We distinguish the cases of massive and
massless shell.\\
a)
$m\neq 0$: 
\begin{align}
&r_{1,2}=-\frac{m^2}{2(M\mp m)}\;,~~~~~
r_{3,4}=\frac{iL}{m}+\frac{ML^2}{m^4}\mp
\frac{\sqrt{2M}L^{5/2}}{m^{9/2}}\e^{-i\pi/4}\notag \;,\\
&r_{5,6}=-\frac{iL}{m}+\frac{ML^2}{m^4}\mp
\frac{\sqrt{2M}L^{5/2}}{m^{9/2}}\e^{i\pi/4}\notag \;.
\end{align}
b) $m= 0$: 
\begin{align}
&r_{1,2}=-\frac{L^{2/3}}{(2M)^{1/3}}\mp \frac{L}{3M}\;,~~~~
r_{3,4}=\frac{L^{2/3}}{(2M)^{1/3}}\e^{i\pi/3}\mp
\frac{L}{3M}\;,~~~r_{5,6}=\frac{L^{2/3}}{(2M)^{1/3}}\e^{-i\pi/3}\mp
\frac{L}{3M}\notag \;. 
\end{align}
All turning points approach zero at $L\to 0$ except for $r_{1,2}$ in
the massive case. Numerically tracing their motion as $L$
decreases from $L_*$,  we find that the physical turning point $A$ of
the reflected 
trajectory is $r_6$ in both cases.

%%%%%%%%%%%%%%%%%%%%%%%%%%%%%%%%%%%%%%%%%%%%%%%%%%%%%%%%%%%%%%%%%%%%%%%%
\section{Gibbons--Hawking terms at the initial-- and final--time
  hypersurfaces}
\label{sec:gibb-hawk-terms}
\begin{figure}
\centerline{\includegraphics[width=6.0cm]{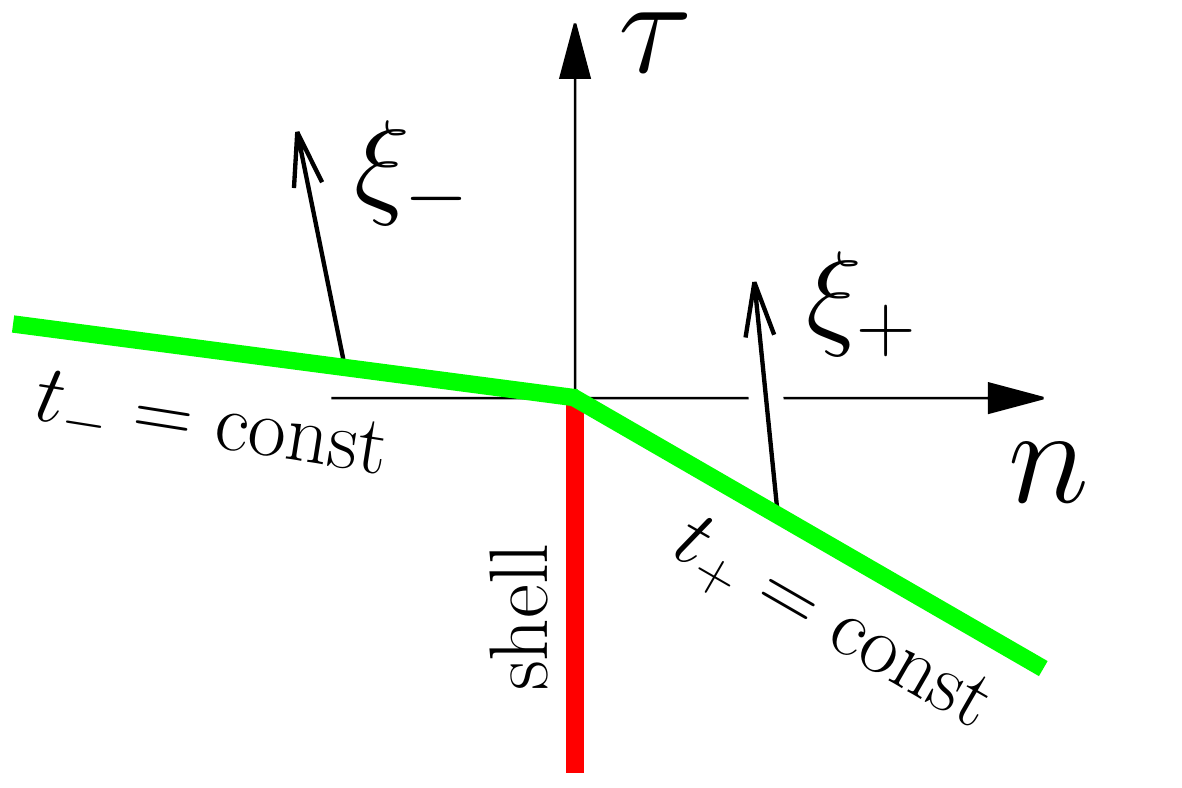}}
\caption{\label{fig:tif}Final--time hypersurface in the Gaussian normal coordinates
  associated with the shell.}
\end{figure}

Since the space--time is almost flat in the beginning and end of the
scattering 
process, one might naively expect that the Gibbons--Hawking terms at $t
_+= t_{i}$ and $t_+ = t_f$ are vanishingly small. However, this
expectation is incorrect. Indeed, it is natural to define the initial
and final hypersurfaces as $t_+ = const$ outside of the shell and 
$t_- = const$ inside it. Since the metric is discontinuous in the
Schwarzschild coordinates, the inner and outer parts of the surfaces
meet at an angle which gives rise to non--zero extrinsic curvature,
see Fig.~\ref{fig:tif}.

For concreteness we focus on the final--time hypersurface. In the
Schwarzschild coordinates the normal vectors to its inner and outer
parts are
\begin{equation}
\label{eq:65}
\xi_-^{\mu} = (1/\sqrt{f_{-}},\, 0,\, 0,\, 0)\;, \qquad\qquad
\xi_+^{\mu} = (1/\sqrt{f_{+}},\, 0,\, 0,\, 0)\;.
\end{equation}
It is easy to see that the extrinsic curvature $K=\nabla_\mu \xi^\mu$
is zero everywhere except for the two--dimensional sphere at the
intersection the hypersurface 
with the shell worldsheet. Let us introduce a Gaussian
normal frame $(\tau,n,\theta^\alpha)$ in the vicinity of the shell,
see Fig.~\ref{fig:tif}. Here $\tau$ is the proper time on the shell,
$n$ is the geodesic distance from it, and $\theta^\alpha$,
$\alpha=2,3$, are the spherical angles. In this frame the metric in
the neighborhood of the shell is essentially flat; corrections due to
nonzero curvature are irrelevant for our discussion.

To find the components of $\xi_+^{\mu}$ and $\xi_-^{\mu}$
in Gaussian normal coordinates,  we project them on $\tau^\mu$ and
$n^\mu$ --- tangent and normal vectors of the shell. The latter in the
inner and outer Schwarzschild coordinates have the form,
\be
\label{eq:taun}
\tau^\mu=\bigg(\frac{\sqrt{\dot r^2+f_\pm}}{f_\pm},\; \dot r,\;
0,\; 0\bigg)~, \qquad\qquad
n^\mu=\bigg(\frac{\dot r}{f_\pm},\; \sqrt{\dot r^2+f_\pm}, \; 0,\;
0\bigg)\;.
\ee
Evaluating the scalar products of (\ref{eq:65}) and (\ref{eq:taun}),
we find,
\be
\label{eq:xipm}
\xi_\pm^\mu=\ch\psi_\pm\, \tau^\mu-\sh\psi_\pm \,n^\mu~,~~~~~
\sh\psi_\pm\equiv\frac{\dot r}{\sqrt{f_\pm}}\;.
\ee
As expected, the normals $\xi_{\pm}^\mu$ do not coincide at the 
position of the shell. To
compute the surface  
integral in the Gibbons--Hawking term, we regularize the
jump by replacing (\ref{eq:xipm}) with  
\be
\label{eq:xi}
\xi^\mu=\ch\psi(n)\, \tau^\mu-\sh\psi(n) \,n^\mu\;,
\ee
where $\psi(n)$ is a smooth function interpolating between $\psi_-$
and $\psi_+$. The expression~(\ref{eq:14}) takes the form,
\begin{equation}
\label{eq:GHtf}
S_{GH}= -\frac{1}{8\pi}\int r^2d^2\theta\int dn\,\frac{ds}{dn}\, K 
=\frac{r^2}{2}(\psi_+-\psi_-)\;,
\end{equation}
where in the second equality we used $ds=dn/\ch\psi$ for the proper
length along the final--time hypersurface and
$K=\d_\m\xi^\m=-\ch\psi\, \psi'$ for its extrinsic curvature.  
Next, we express $\psi_\pm(r)$ from the shell equation of motion
(\ref{eq:30}) and expand Eq.~(\ref{eq:GHtf}) at large $r$. Keeping only
non--vanishing terms at $r=r_f\to + \infty$, we obtain Eq.~(\ref{eq:68}) for
the final--time Gibbons--Hawking term.   

For the initial--time hypersurface the derivation is the same, 
the only difference is in the sign of $\xi^\mu$ which is now
past--directed. However, this is compensated by the change of sign of
$\dot{r}$. 
One concludes that  the 
Gibbons--Hawking term at $t_+ = t_i$ is obtained from the one 
at $t_+=t_f$ by the substitution $r_f \to r_i$. 

Note that expression (\ref{eq:GHtf}) is valid also in the model of
Sec.~\ref{sec:massless-shell-ads_4} describing massless shell in
AdS. It is straightforward to see that in the latter case the
Gibbons--Hawking terms vanish at $r_{i,f}\to\infty$  due to growth
of the metric functions (\ref{eq:60}) at large $r$.   

%%%%%%%%%%%%%%%%%%%%%%%%%%%%%%%%%%%%%%%%%%%%%%%%%%%%%%%%%%%%%%%%%%%%%%%%
\section{Shell  self--gravity at order $1/r$}
\label{app:free}
Let us  construct the action for a  neutral 
shell in asymptotically flat space--time taking into account
its self--gravity at order $1/r$. To this end we recall that  the
shell is assembled from  particles of mass $\delta m$, see
Appendix~\ref{sec:shell-rotating-dust}. Every 
particle moves in the mean field of other particles. Thus,  a
new particle added to the shell changes the
action of the system\footnote{Angular motion of the particle gives
  $1/r^{2}$ contributions to the Lagrangian which are irrelevant in
  our approximation.} by
\begin{equation}
\label{eq:74}
\delta S = - \int \delta m \,  d\tau
=\int dt_+\bigg(-\delta m\sqrt{1-v^2}+\frac{\delta m
  (1+v^2)}{\sqrt{1-v^2}}\,
\frac{\bar M}{r}\bigg)
\;,
\end{equation}
where $v=dr/dt_+$ is the shell velocity in the asymptotic coordinates,
$\bar M$ is its energy, and we expanded the
proper time $d\tau$ up to the  
first order in $1/r$ in the second equality. At
the leading order in $1/r$,
\be
\bar M=\frac{\bar m}{\sqrt{1-v^2}}\;,
\ee
where $\bar m$ is the shell mass before adding the particle. Now, we
integrate Eq.~(\ref{eq:74}) from $\bar m=0$ to the actual shell mass
$m$ and obtain the desired action, 
\begin{equation}
\label{eq:S0}
S_0 = \int dt_+\bigg(-m\sqrt{1-v^2}+\frac{m^2
  (1+v^2)}{2r(1-v^2)}
\bigg)
\;.
\end{equation}
From this expression one reads off the canonical momentum and energy
of the shell,
\begin{align}
\label{eq:75}
&p = \frac{m v}{\sqrt{1-v^2}}+\frac{2m^2 v}{r(1-v^2)^2}\;,\\
\label{eq:751}
&M=\frac{m}{\sqrt{1-v^2}}+\frac{m^2(-1+4v^2+v^4)}{2r(1-v^2)^2}\;.
\end{align}
Expressing the shell velocity from Eq.~(\ref{eq:751}) and 
substituting\footnote{In this calculation the $1/r$ terms are treated as corrections.} it into
Eq.~(\ref{eq:75}), we obtain Eq.~(\ref{eq:77}) from the main text. 

%%%%%%%%%%%%%%%%%%%%%%%%%%%%%%%%%%%%%%%%%%%%%%%%%%%%%%%%%%%%%%%%%%%%%%%%

\end{document}